\documentclass[a4paper,11pt]{article}

\pdfoutput=1

\usepackage{amsmath,amssymb,mathtools}
\usepackage{color}
\usepackage{graphicx}
\usepackage{subfigure}
\usepackage{cite}
\usepackage[colorlinks=true,linkcolor=blue, citecolor=red, urlcolor=blue, bookmarks]{hyperref}
\usepackage{multirow,makecell}
\usepackage{textcomp}
\usepackage{wasysym}

\usepackage[text={17.1cm,24.6cm},centering]{geometry} %% thanks to Chao Wu

\numberwithin{equation}{section}

\def \be {\begin{equation}}
\def \ee {\end{equation}}
\def \ba {\begin{array}}
\def \ea {\end{array}}
\def \bea{\begin{eqnarray}}
\def \eea{\end{eqnarray}}
\def \nn {\nonumber}

\def \d {\delta}

\def \ve {\varepsilon}
\def \m {\mu}
\def \n {\nu}

\def \s {\sigma}

\def \r {\rho}

\def \th {\theta}

\def \vph {\varphi}
\def \vphi {\varphi}
\def \t {\tau}
\def \z {\zeta}

\def \cF {\mathcal F}

\def \rC {\mathrm C}
\def \rD {\mathrm D}

\def \rZ {\mathrm Z}

\def \p {\partial}

\def \f {\frac}

\def \lra {\leftrightarrow}
\def \sr {\sqrt}
\def \td {\tilde}

\def \inf {\infty}

\def \lag {\langle}
\def \rag {\rangle}

\def \ep {\mathrm{e}}
\def \ii {\mathrm{i}}

\def \Re {{\textrm{Re}}}
\def \Im {{\textrm{Im}}}
\def \tr {\textrm{tr}}

\def \diag {{\textrm{diag}}}

\def \and {{~\textrm{and}~}}

\begin{document}

\title{
\textbf{Excited state R\'enyi entropy and subsystem distance in two-dimensional non-compact bosonic theory\\I. Single-particle states
}
%\textbf{Excited state R\'enyi entropy and subsystem distance in harmonic chain and two-dimensional free boson theory}
%\textbf{Excited state R\'enyi entropy and Schatten distance in gapless harmonic chain and 2D free massless boson theory}
%\\or\\
%\textbf{The success and failure of the lattice Bisognano-Wichmann modular Hamiltonian}\\or\\
%\blue{We need more exiting title!}
}
\author{
Jiaju Zhang$^{1}$ %\footnote{jzhang@sissa.it},
and
M. A. Rajabpour$^{2}$%\footnote{mohammadali.rajabpour@gmail.com}
}
\date{}
\maketitle
\vspace{-12mm}
\begin{center}
{\it
$^{1}$SISSA and INFN, Via Bonomea 265, 34136 Trieste, Italy\\\vspace{1mm}
%$^{1}$Scuola Internazionale Superiore di Studi Avanzati (SISSA), Via Bonomea 265, 34136 Trieste, Italy\\\vspace{1mm}
%$^{2}$INFN Sezione di Trieste, Via Bonomea 265, 34136 Trieste, Italy\\\vspace{1mm}
%$^{3}$International Centre for Theoretical Physics (ICTP), Strada Costiera 11, 34151 Trieste, Italy\\\vspace{1mm}
$^{2}$Instituto de Fisica, Universidade Federal Fluminense,\\
Av. Gal. Milton Tavares de Souza s/n, Gragoat\'a, 24210-346, Niter\'oi, RJ, Brazil
}
\vspace{10mm}
\end{center}

\begin{abstract}
  We investigate the R\'enyi entropy of the excited states produced by the current and its derivatives in the two-dimensional free massless non-compact bosonic theory, which is a two-dimensional conformal field theory. We also study the subsystem Schatten distance between these states. The two-dimensional free massless non-compact bosonic theory is the continuum limit of the finite periodic gapless harmonic chains with the local interactions. We identify the excited states produced by current and its derivatives in the massless bosonic theory as the single-particle excited states in the gapless harmonic chain. We calculate analytically the second R\'enyi entropy and the second Schatten distance in the massless bosonic theory. We then use the wave functions of the excited states and calculate the second R\'enyi entropy and the second Schatten distance in the gapless limit of the harmonic chain, which match perfectly with the analytical results in the massless bosonic theory. We verify that in the large momentum limit the single-particle state R\'enyi entropy takes a universal form. We also show that in the limit of large momenta and large momentum difference the subsystem Schatten distance takes a universal form but it is replaced by a new corrected form when the momentum difference is small. Finally we also comment on the mutual R\'enyi entropy of two disjoint intervals in the excited states of the two-dimensional free non-compact bosonic theory.
\end{abstract}

\baselineskip 18pt
\thispagestyle{empty}
\newpage

\tableofcontents

%\begin{center}
%\textbf{\LARGE Excited state R\'enyi entropy and subsystem distance in harmonic chain}\\~\\
%\today
%\end{center}

\section{Introduction}

The low energy physics of many one-dimensional critical quantum chains in the continuum limit can be described by two-dimensional (2D) conformal field theories (CFTs) \cite{henkel1999conformal}.
It is interesting to compare various quantities in CFTs with those in the corresponding critical quantum chains.
One important quantity is the entanglement entropy, which plays a key role in better understanding of the quantum many-body systems and the quantum field theories (QFTs) \cite{Amico:2007ag,Eisert:2008ur,calabrese2009entanglement,Laflorencie:2015eck,Witten:2018lha}.
To calculate the entanglement entropy, one first divides the total system with the density matrix $\r$ into the subsystem $A$ and its complement $B$, and then traces out the degrees of freedom of $B$, to get the reduced density matrix (RDM) $\r_A=\tr_B\r$. Then the entanglement entropy is just the von Neumann entropy
\be
S_A = - \tr_A( \r_A \log \r_A ).
\ee
The entanglement entropy is often calculated as the $n\to1$ limit of the R\'enyi entropy
\be
S_A^{(n)} = -\f{1}{n-1} \log \tr_A\r_A^n,
\ee
where $n$ can be any positive real number.
The entanglement entropy and in general the R\'enyi entropy of one single interval in the ground state of various 2D CFTs and quantum chains have been investigated in full detail in the last three decades \cite{Bombelli:1986rw,Srednicki:1993im,Callan:1994py,Holzhey:1994we,peschel1999density,Peschel:1999DensityMatrices,%
Chung2000Densitymatrix,chung2001density,Vidal:2002rm,peschel2003calculation,Latorre:2003kg,jin2004quantum,Korepin:2004zz,Plenio:2004he,Calabrese:2004eu,%
Cramer:2005mx,Casini:2005rm,Casini:2005zv,Casini:2009sr,Calabrese:2009qy,peschel2009reduced,peschel2012special}.
Especially, the R\'enyi entropy of a length $\ell$ interval on a one-dimensional infinity gapless system in the ground state takes the logarithmic formula  \cite{Holzhey:1994we,Vidal:2002rm,jin2004quantum,Korepin:2004zz,Calabrese:2004eu}
\be
S_{A,G}^{(n)} = \f{c(n+1)}{6n} \log \ell + c_n,
\ee
with the universal central charge $c$ and the non-universal constant $c_n$.
There are also many studies regarding multi-interval entanglement entropy and R\'enyi entropy in the ground state \cite{Casini:2004bw,Furukawa:2008uk,Casini:2008wt,Facchi:2008Entanglement,Caraglio:2008pk,Casini:2009vk,Calabrese:2009ez,Alba:2009ek,%
Igloi:2009On,Fagotti:2010yr,Headrick:2010zt,Calabrese:2010he,Alba:2011fu,Rajabpour:2011pt,Coser:2013qda,DeNobili:2015dla,Coser:2015dvp,%
Dupic:2017hpb,Ruggiero:2018hyl,Arias:2018tmw}.
Finally for the study of single-interval entanglement of the excited states in QFTs and quantum chains see
\cite{Alba:2009th,Alcaraz:2011tn,Berganza:2011mh,Pizorn2012Universality,Essler2013ShellFilling,Berkovits2013Twoparticle,Taddia:2013Entanglement,%
Storms2014Entanglement,Palmai:2014jqa,Calabrese:2014Entanglement,Molter2014Bound,Taddia:2016dbm,Castro-Alvaredo:2018dja,Castro-Alvaredo:2018bij,Murciano:2018cfp,%
Castro-Alvaredo:2019irt,Castro-Alvaredo:2019lmj,Jafarizadeh:2019xxc,Capizzi:2020jed,You:2020osa,Haque:2020Entanglement,Wybo:2020fiz}.

In this paper, we investigate the R\'enyi entropy in the excited states produced by the current and its derivatives in the 2D free massless non-compact bosonic theory, which is a 2D CFT with central charge $c=1$, and the corresponding quantity in the gapless limit of the harmonic chain.
Most of the previous works were focused on the R\'enyi entropy of the ground state of the 2D free bosonic theory and the harmonic chain \cite{Bombelli:1986rw,Srednicki:1993im,Callan:1994py,Holzhey:1994we,Peschel:1999DensityMatrices,Chung2000Densitymatrix,%
peschel2003calculation,Calabrese:2004eu,Plenio:2004he,Cramer:2005mx,Furukawa:2008uk,Casini:2009sr,Calabrese:2009qy,%
peschel2009reduced,Calabrese:2009ez,Headrick:2010zt,peschel2012special,Coser:2013qda,DeNobili:2015dla}.
We note that the R\'enyi entropy of the low-lying excited states of the 2D free massless compact bosonic theory has been already calculated and compared with the  numerical results coming from the spin-1/2 XX chain in \cite{Alcaraz:2011tn,Berganza:2011mh,Taddia:2016dbm}.
The recent investigations on the excited state R\'enyi entropy of the bosonic theory and harmonic chain were mainly focused on the gapped regime \cite{Castro-Alvaredo:2018dja,Castro-Alvaredo:2018bij,Castro-Alvaredo:2019irt}.
In this paper, we will calculate the second R\'enyi entropy in the gapless regime of the bosonic theory and harmonic chain.
We identify the excited states of current and its derivatives in the free massless non-compact bosonic theory with the single-particle states in the gapless harmonic chain.
We calculate the second R\'enyi entropy of the single-particle excited states in the gapless limit of the harmonic chain, using the mini version of the wave function method elaborated in \cite{Castro-Alvaredo:2018dja,Castro-Alvaredo:2018bij}.
We compare the analytical CFT results and the numerical lattice results, and find perfect matches.

In quantum information theory, it is often important to know quantitatively the difference between two density matrices, especially for the subsystem RDMs.
Consequently the concept has been used and calculated in different areas.
The subsystem distance was used in \cite{fagotti2013reduced} to characterize the thermalization of subsystems after a global quench \cite{Calabrese:2005in,Calabrese:2006rx,Calabrese:2007mtj}.
The subsystem distance of the low-lying states in the 2D free massless fermionic and compact bosonic theories were already calculated recently and compared with the results coming from the critical Ising chain and XX chain in \cite{Zhang:2019wqo,Zhang:2019itb}.
In \cite{Mendes-Santos:2019tmf,Zhang:2020mjv} the subsystem distance was used to quantify the precision of the approximate entanglement Hamiltonian coming from the discretization of the Bisognano-Wichmann modular Hamiltonian in critical quantum spin chains \cite{Dalmonte:2017bzm,Giudici:2018izb}.
The subsystem distance was also used in \cite{Zhang:2019kwu} to characterize the local operator quench in 2D CFTs and spin chains \cite{Nozaki:2014hna,Nozaki:2014uaa,He:2014mwa}.
Recently, the subsystem distance in the thermal states of the finite size critical XY chains was investigated in \cite{Arias:2020sgz}.
There are many definitions of the distance between two states, see for example \cite{nielsen2010quantum,hayashi2017quantum,watrous2018theory}. In this paper we will use the Schatten distance between the RDMs $\r_A$ and $\s_A$ normalized as
\be
D_n( \r_A,\s_A ) = \Big( \f{\tr_A| \r_A - \s_A |^n}{2\tr_A\r_{A,G}^n} \Big)^{1/n},
\ee
where we use the ground state RDM $\r_{A,G}$ to cancel the UV divergence as in \cite{Zhang:2019itb}. The $n=1$ case of the Schatten distance is the trace distance $D( \r_A,\s_A ) = \f12 \tr_A| \r_A - \s_A |$, which we will not consider in the current paper.
In this paper we will calculate the second Schatten distance between the ground state and the excited states of the current and its derivatives in the 2D free massless bosonic theory and compare with the ones between the ground state and the single-particle excited states in the gapless limit of the harmonic chain.

The paper is organized as follows:
In Section~\ref{secHC} we review the basic properties of the ground state and the single-particle states and their wave functions in the harmonic chain with the local couplings.
In Section~\ref{secID} we elaborate the identification of the excited states of the current and its derivatives in the 2D free massless non-compact bosonic theory with the single-particle states in the gapless harmonic chain.
We calculate the second single-interval R\'enyi entropy analytically in the 2D free massless non-compact bosonic theory and numerically in the gapless limit of the harmonic chain and compare the results in Section~\ref{secRenyi}.
We do the same for the second Schatten distance in Section~\ref{secSchatten}.
We consider the R\'enyi mutual information of two disjoint intervals in Section~\ref{secRMI}.
We conclude with discussions in Section~\ref{secCon}.
We collect the CFT results of the R\'enyi entanglement entropy and Schatten distances  in Appendices~\ref{appRenyi} and \ref{appSchatten} respectively.

\section{Harmonic chain basics: ground and single-particle states}\label{secHC}

In this section we review the textbook properties of the discrete version of the 2D free massive bosonic theory, i.e. the harmonic chains with the local couplings, which will help us to fix the notation.
We consider the 2D free non-compact bosonic theory with the Lagrangian density
\be
\mathcal L = - \f{1}{8\pi} ( \eta^{\m\n} \p_\m\phi\p_\n\phi+m^2\phi^2 ),
\ee
with the metric $\eta^{\m\n}=\diag(-1,1)$, derivatives $\p_\m=(\p_t,\p_u)$, real temporal coordinate $t$, spatial coordinate $u$, and the mass (or equivalently gap) $m$.
The Hamiltonian density is
\be
\mathcal H = \f{1}{8\pi} [ (\p_t\phi)^2 + (\p_u\phi)^2 + m^2\phi^2 ].
\ee
The discrete version of 2D free non-compact bosonic theory is just the harmonic chain with the local couplings%
\footnote{We note that many of our formulas can be applied without any changes to also  more general harmonic chains such as
\[ H = \f{1}{2} \sum_{j_1,j_2=1}^L \big( M_{j_1j_2} p_{j_1} p_{j_2}+ N_{j_1j_2} q_{j_1} q_{j_2} \big), \]
with $L\times L$ real symmetric coupling matrices $M$, $N$.}
\be \label{HCwNI}
H = \f{1}{2} \sum_{j=1}^L \big[ p_j^2 + m^2 q_j^2 + (q_j-q_{j+1})^2 \big].
\ee
Here we consider $L$, the size of the full system, an even integer and impose the periodic boundary condition $q_{L+1}=q_1$. The operators $q_j$, $p_j$ satisfy the canonical commutation relations
\be
[q_{j_1},q_{j_2}] = [p_{j_1},p_{j_2}] = 0, ~~ [q_{j_1},p_{j_2}] = \ii \d_{j_1j_2}.
\ee
The model suffers from IR divergence in the gapless limit $m\to0$, so we need to keep the gap $m$ general in the calculations and take the small $m$ limit at the end.

To diagonalize the Hamiltonian, one can make the Fourier transformation
\be
q_j = \f{1}{\sr{L}} \sum_k \ep^{-\f{2\pi\ii j k}{L}}\vph_k, ~~
p_j = \f{1}{\sr{L}} \sum_k \ep^{-\f{2\pi\ii j k}{L}}\pi_k,
\ee
with the integer momenta
\be
k=1-\f{L}{2},\cdots,-1,0,1,\cdots,\f{L}{2}-1,\f{L}{2}.
\ee
The Hamiltonian becomes
\be
H = \f12 \sum_k ( \pi_k^\dag \pi_k + \ve_k^2\vph_k^\dag\vph_k ),
\ee
with the spectrum
\be
\ve_k = \sr{m^2+4\sin^2\f{\pi k}{L}}.
\ee
Note that $\vph_k^\dag=\vph_{-k}$, $\pi_k^\dag=\pi_{-k}$.
One can then define the ladder operators
\be
b_k=\sr{\f{\ve_k}{2}}\Big( \vph_k + \f{\ii}{\ve_k} \pi_k \Big), ~~
b_k^\dag=\sr{\f{\ve_k}{2}}\Big( \vph_k^\dag - \f{\ii}{\ve_k} \pi_k^\dag \Big),
\ee
satisfying
\be
[b_{k_1},b_{k_2}] = [b_{k_1}^\dag,b_{k_2}^\dag] = 0, ~~ [b_{k_1},b_{k_2}^\dag] = \d_{k_1k_2}.
\ee
The Hamiltonian becomes
\be
H = \sum_k \ve_k \Big( b_k^\dag b_k +\f12 \Big).
\ee

The ground state $|G\rag$ is annihilated by all the lowering operators
\be
b_k | G \rag = 0, ~ k=1-\f{L}{2},\cdots,\f{L}{2}.
\ee
The ground state wave function in the coordinate basis is
\be \label{WFQG}
\lag Q | G \rag = \Big( \det\f{W}{\pi} \Big)^{1/4} \ep^{ -\f12 Q^T W Q },
\ee
where the coordinates $Q=(q_1,\cdots,q_L)$ and the $L\times L$ real symmetric matrix
\be
W_{j_1j_2} = \f1{L}\sum_k\ve_k\cos\f{2\pi k(j_1-j_2)}{L}.
\ee
The inverse matrix can be also calculated easily as
\be
W_{j_1j_2}^{-1} = \f1{L}\sum_k\f{1}{\ve_k}\cos\f{2\pi k(j_1-j_2)}{L}.
\ee
The density matrix of the total system is
\be \label{rhoG}
\lag Q | \r_G | Q' \rag = \sr{\det\f{W}{\pi}}\ep^{-\f12 Q^T W Q-\f12 Q'^T W Q'}.
\ee

The energy eigenstates can be obtained by applying the raising operators on the ground state. In this paper we only consider the states with the excitation of only one quasiparticle
\be
|k\rag = b_k^\dag | G \rag,
\ee
which we call the single-particle states.
The wave function of the single-particle state $|k\rag$ is
\be
\lag Q | k \rag = \lag Q | G \rag Q^T v_k,
\ee
with the ground state wave function (\ref{WFQG}) and the vector components
\be
[v_{k}]_j = \sr{\f{2\ve_k}{L}} \ep^{-\f{2\pi\ii j k}{L}}, ~ j=1,2,\cdots,L.
\ee
One can easily check that these states are already in orthonormal basis.
Finally the density matrix of the total system  for the single-particle state is
\be\label{rhok}
\lag Q | \r_k | Q' \rag = \lag Q | \r_G | Q' \rag Q^T V_k Q',
\ee
where $\lag Q | \r_G | Q' \rag$ is the ground state density matrix (\ref{rhoG}) and  $V_k = v_k v_k^\dag$ is an $L\times L$ hermitian matrix.

\section{Identification of CFT and harmonic chain states}\label{secID}

In this section, we elaborate the identification of the excited states of current and its derivatives in the 2D free massless non-compact bosonic theory with the single-particle excited states in the gapless harmonic chain.
The 2D free massless non-compact bosonic theory, which is a 2D CFT with the central charge $c=1$, is the continuum limit of the gapless harmonic chain with the local couplings (\ref{HCwNI}).
We follow mainly \cite{Zhang:2019kwu}, however one can also see \cite{Itoyama:1986ad,Itoyama:1988zn,Koo:1993wz,Zou:2019dnc,Zou:2019iwr} for more rigorous identifications of the states and operators in the 2D CFTs and critical lattices.
One can consult \cite{DiFrancesco:1997nk,Blumenhagen:2009zz} for the basics of the 2D free massless bosonic theory.

We consider the 2D free massless non-compact bosonic theory on a cylinder with complex coordinates
\be
w=u-\ii\t=u+t, ~~ \bar w=u+\ii\t=u-t.
\ee
Note that $u$ is the spatial coordinate, $\t$ is the Euclidean time, and $t$ is the real time. In the spatial direction we have $u\simeq u+L$.
The scalar field $\phi$ can be written as a sum of the holomorphic and anti-holomorphic parts
\be
\phi(u,t) = \vphi(u+t) + \bar \vphi(u-t).
\ee
Then the current operators are
\be
J(w) = \ii \p\vphi(w), ~~ \bar J(\bar w) = \ii \bar\p\bar\vphi(\bar w),
\ee
which are primary operators with conformal weights (1,0) and (0,1) respectively.
At fixed time $t=0$, we have
\be
\ii\p_u\phi(u,0) = J(u) + \bar J(u).
\ee
The cylinder with coordinate $w$ can be mapped to a plane with coordinate $z$ by the transformation
\be
z=\ep^{\f{2\pi\ii w}{L}}. %, ~~ \bar z=\ep^{-\f{2\pi\ii \bar w}{L}}.
\ee
The current operator transforms as
\be
J(w)=\f{\p z}{\p w} J(z). %, ~~ \bar J(\bar w)=\f{\bar\p \bar z}{\bar\p \bar w} \bar J(\bar z).
\ee
On the plane there is mode expansion
\be
J(z) = \sum_{k\in\rZ} \f{J_k}{z^{k+1}}.
\ee
Note that
\be
J_k |G\rag =0, ~ k>-1.
\ee
At $t=0$ we get the current operator applied on the ground state on the cylinder
\be
J(u)|G\rag = \f{2\pi\ii}{L} \sum_{k=1}^{+\inf} \ep^{\f{2\pi\ii k u}{L}}J_{-k}|G\rag.
\ee
Similarly, we get
\be
\bar J(u)|G\rag = -\f{2\pi\ii}{L} \sum_{k=1}^{+\inf} \ep^{-\f{2\pi\ii k u}{L}}\bar J_{-k}|G\rag.
\ee
Finally, we obtain
\be \label{expoCFT}
\ii\p_u\phi(u,0) |G\rag = \f{2\pi\ii}{L} \sum_{k=1}^{+\inf}
                       \Big( \ep^{\f{2\pi\ii k u}{L}} J_{-k}|G\rag
                           - \ep^{-\f{2\pi\ii k u}{L}}\bar J_{-k}|G\rag \Big).
\ee
Note that for $k>0$
\be
J_{-k}|G\rag = \f{\p^{k-1}J(0)}{(k-1)!}|G\rag,
\ee
and it is normalized as $\lag G|J_k J_{-k}|G\rag=k$.
 Similar formula is valid for the state $\bar J_{-k}|G\rag$.

There is a simple correspondence between 2D free massless non-compact bosonic theory and the gapless harmonic chain
\bea
\phi(u,0) &\lra& \sr{4\pi} q_j , \nn\\
\ii \p_u \phi(u,0) &\lra& \ii \sr{4\pi} ( q_{j+1} - q_j ) .
\eea
We take the gapless limit $m\to0$ and continuum limit $L\to+\inf$ of the lattice and only consider the low-lying excited states, and this allows us to write
\be
\sin\f{\pi k}{L} \to \f{\pi k}{L}.
\ee
In the harmonic chain we get
\be \label{expoLattice}
\ii \sr{4\pi} ( q_{j+1} - q_j ) |G\rag = - \f{2\pi}{L} \sum_{k=1}^{+\inf}
\Big( \ep^{\f{2\pi\ii k(j+1/2)}{L}} \sr{k} |k\rag
     -\ep^{-\f{2\pi\ii k(j+1/2)}{L}} \sr{k} |{-}k\rag \Big).
\ee
Note that $|k\rag=b_k^\dag|G\rag$, $|{-}k\rag=b_{-k}^\dag|G\rag$.
Here we do not care about the overall normalizations or the phases of the states.
Comparing the CFT and lattice expressions (\ref{expoCFT}) and (\ref{expoLattice}), we identify the states in the 2D free massless bosonic theory and the gapless harmonic chain and with $k>0$
\bea \label{identification}
\f{J_{-k}}{\sr{k}} |G\rag = \f{\p^{k-1}J(0)}{\sr{k!(k-1)!}} |G\rag &\lra& |k\rag, \nn\\
\f{\bar J_{-k}}{\sr{k}}|G\rag = \f{\bar\p^{k-1}\bar J(0)}{\sr{k!(k-1)!}} |G\rag &\lra& |{-}k\rag.
\eea
In summary, we have shown that the excited states of current and its derivatives in the 2D free massless non-compact bosonic theory are the same as the single-particle states in the scaling limit of the gapless harmonic chain.

\section{R\'enyi entropy}\label{secRenyi}

In this section we consider the R\'enyi entropy of a single interval $A$ of length $\ell$ on a circle of length $L$. It is convenient to define the ratio $x=\f{\ell}{L}$.
We first calculate the R\'enyi entropy analytically in the 2D free massless non-compact bosonic theory and numerically in the harmonic chain, and then compare the two results.

\subsection{Massless bosonic theory}

In the 2D free massless non-compact bosonic theory we first consider the R\'enyi entropy of an interval $A=[0,\ell]$ on a circle of length $L$ in the ground state.
In a 2D CFT with central charge $c=1$, one expects the universal R\'enyi entropy\cite{Holzhey:1994we,Calabrese:2004eu,Calabrese:2009qy}
\be
S_{A,G}^{(n)} = \f{n+1}{6n} \log\Big(\f{L}{\pi}\sin\f{\pi\ell}{L}\Big) + c_n,
\ee
with non-universal constant $c_n$.
However, the R\'enyi entropy suffers from IR divergence in the massless limit $m\to0$, and the single-interval R\'enyi entropy on an infinite line is \cite{Casini:2005zv,Casini:2009sr}
\be
S_{A,G}^{(n)} = \f{n+1}{6n} \log \ell + \f12\log\log\f{1}{m\ell} + c_n'.
\ee
We expect a similar form for the single-interval R\'enyi entropy on a circle, i.e. the sum of the universal CFT term, the IR divergent term, and the constant term.
The possible IR divergent terms are $\f12\log\log\f{1}{m\ell}$, $\f12\log\log\f{1}{m L}$, $\f12\log\f{1}{m\ell}$, $\f12\log\f{1}{m L}$, and we were able to check each of them against the numerical results.
By guessing and numerical fitting in the next subsection we get the second R\'enyi entropy of a length $\ell$ interval on a length $L$ circle in the massless limit of the 2D free non-compact bosonic theory\footnote{In \cite{Yazdi:2016cxn} there is a similar result
\[
S_{A,G}^{(2)} = \f14 \log\Big(\f{L}{\pi}\sin\f{\pi\ell}{L}\Big) + \f12 \log\f{1}{m \ell} + s'_2,
\]
where the ``constant'' $s'_2=s_2+\f12\log\f{\ell}{L}$ would actually depend on the ratio $x=\f{\ell}{L}$.}
\be \label{SAG2}
S_{A,G}^{(2)} = \f14 \log\Big(\f{L}{\pi}\sin\f{\pi\ell}{L}\Big) + \f12 \log\f{1}{m L} + s_2.
\ee
The constant $s_2$ is independent of $m$, $L$, $\ell$ in the massless and continuum limit.

As we showed in Section~\ref{secID}, there is a one-to-one correspondence between the states in the 2D free massless bosonic theory with the single-particle excited states in the gapless harmonic chain as (\ref{identification}) with $k>0$.
In the massless bosonic theory, for $r=0,1,\cdots$ we construct the density matrices of the total system
\bea
&& \r_{\p^r J} = \f{1}{r!(r+1)!} {\p^rJ(0)} | G \rag \lag G | {\p^rJ(\inf)} , \nn\\
&& \r_{\bar\p^r \bar J} = \f{1}{r!(r+1)!} {\bar\p^r\bar J(0)} | G \rag \lag G | {\bar\p^r\bar J(\inf)},
\eea
from which we construct the RDMs $\r_{A,\p^r J}$, $\r_{A,\bar\p^r \bar J}$.

As \cite{Alcaraz:2011tn,Berganza:2011mh}, we use $\cF_{A,X}^{(n)}$ to denote the difference between the R\'enyi entropy $S_{A,X}^{(n)}$ in the excited state $|X\rag$ and the ground state R\'enyi entropy $S_{A,G}^{(n)}$ as
\be
S_{A,X}^{(n)} = S_{A,G}^{(n)} - \f{1}{n-1} \log \cF_{A,X}^{(n)}.
\ee
More explicitly, we have
\be
\cF_{A,X}^{(n)} = \f{\tr_A\r_{A,X}^n}{\tr_A\r_{A,G}^n}.
\ee
Following \cite{Alcaraz:2011tn,Berganza:2011mh,Taddia:2016dbm}, especially \cite{Taddia:2016dbm}, we write the second single-interval R\'enyi entropy on a cylinder as a two-point function on a two-fold plane
\bea \label{fourpointfunction1}
&& \cF_{A,\p^r J}^{(2)} = \f{1}{[r!(r+1)!]^2} \lag {\p^rJ(0_1)} {\p^rJ(\inf_1)} {\p^rJ(0_2)} {\p^rJ(\inf_2)} \rag_{\rC^2}, \nn\\
&& \cF_{A,\bar \p^r \bar J}^{(2)} = \f{1}{[r!(r+1)!]^2} \lag {\bar \p^r\bar J(0_1)} {\bar \p^r\bar J(\inf_1)} {\bar \p^r\bar J(0_2)} {\bar \p^r\bar J(\inf_2)} \rag_{\rC^2}.
\eea
The subscripts $1$, $2$ of the coordinates $0_1$, $\inf_1$, $0_2$, $\inf_2$ are the replica indices.
On each replica of two-fold plane there are two operators inserted, one at the origin and another at the infinity.
The two replicas are connected along the cut $[\ep^{-\pi\ii\ell/L},\ep^{\pi\ii\ell/L}]$.
The two-fold plane with coordinate $z$ can be mapped to a plane with coordinate $\z$ through the conformal transformation
\be
\z(z) = \Big( \f{\ep^{-\pi\ii\ell/L} z-1}{z-\ep^{-\pi\ii\ell/L}} \Big)^{1/2}.
\ee
Evaluating the four-point function on the plane, we get the excited state R\'enyi entropy in the CFT.
Explicitly, with $x=\f{\ell}{L}$ we obtain
\be \label{cFAprJ2}
\cF_{A,J}^{(2)} = \frac{1}{128} [ 99 + 28 \cos (2 \pi  x)+\cos (4 \pi  x) ],
\ee
which has been derived in \cite{Alcaraz:2011tn,Berganza:2011mh,Essler2013ShellFilling,Calabrese:2014Entanglement}.
We also obtain new higher level results that are shown in Appendix~\ref{appRenyi}.
It is easy to see $\cF_{A,\p^r J}^{(2)}=\cF_{A,\bar\p^r \bar J}^{(2)}$ in CFT, parallel to $\cF_{A,k}^{(2)}=\cF_{A,-k}^{(2)}$ in the harmonic chain.

\begin{figure}[t]
  \centering
  % Requires \usepackage{graphicx}
  \includegraphics[height=0.3\textwidth]{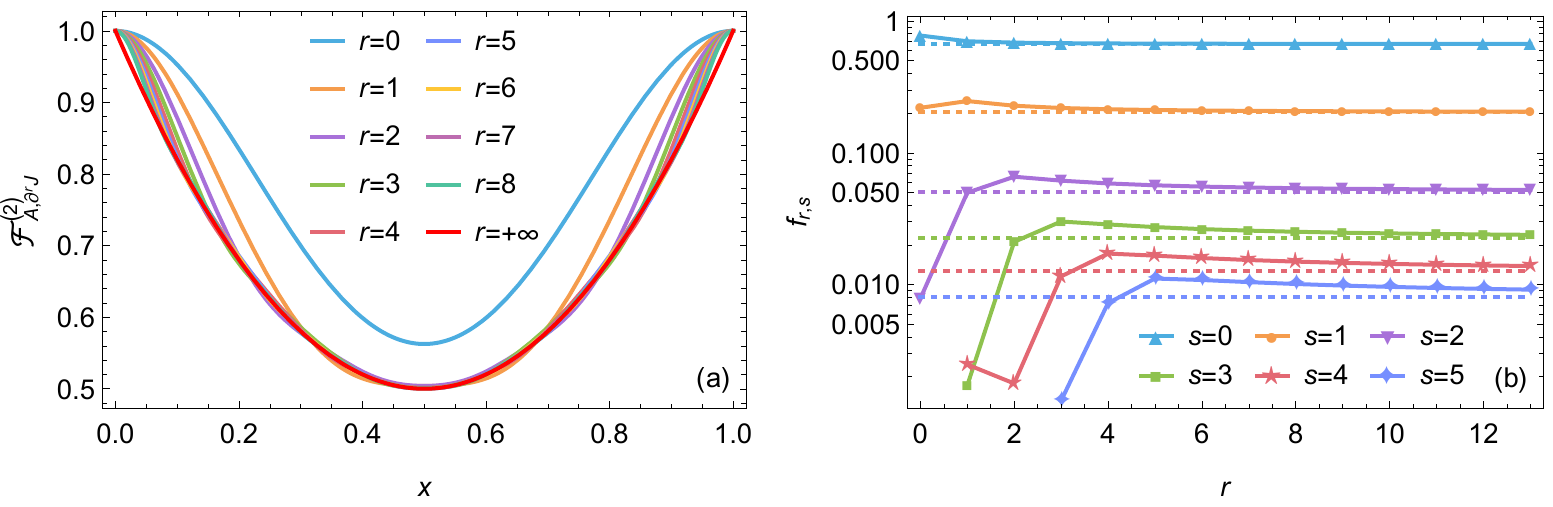}\\
  \caption{The single-interval R\'enyi entropy of the excited sates of the current and its derivatives in the 2D free massless bosonic theory (left) and the coefficients (\ref{fourier1}) in the Fourier expansion (right).
  The dashed lines in the right panel are the Fourier coefficients (\ref{fourier2}) for the conjectured result at $r=+\inf$ (\ref{univRenyiCFT}).
  We verify (\ref{univRenyiCFT}).}\label{RenyiCFT}
\end{figure}

In the left panel of Fig.~\ref{RenyiCFT}, we plot the CFT results $\cF_{A,\p^r J}^{(2)}$ with $r=0,1,\cdots,8$, and it indicates that $\cF_{A,\p^r J}^{(2)}$ approaches to a $r$-independent function in the $r\to+\inf$ limit
\be \label{univRenyiCFT}
\lim_{r \to +\inf} \cF_{A,\p^r J} ^{(2)} = 1-2x+2x^2.
\ee
This is just the universal R\'enyi entropy in the single-particle state studied in \cite{Castro-Alvaredo:2018dja,Castro-Alvaredo:2018bij,Castro-Alvaredo:2019irt,Castro-Alvaredo:2019lmj}.
In the single-particle state $\r_k$, the quasiparticle has the probability $x$ inside the subsystem $A$ and probability $1-x$ outside of $A$.
For $A$, one could write the state with one quasiparticle as $|1]$ and the state with no quasiparticle as $|0]$. Then the RDM is just \cite{Castro-Alvaredo:2018dja,Castro-Alvaredo:2018bij,Castro-Alvaredo:2019irt,Castro-Alvaredo:2019lmj}
\be \label{rAkx111mx00}
\r_{A,k} = x |1] [1| + (1-x) |0] [0|,
\ee
and it leads to easily the R\'enyi entropy (\ref{univRenyiCFT}).
One can write the CFT result (\ref{cFAprJ2}) and the ones in Appendix~\ref{appRenyi} as
\be \label{fourier1}
\cF_{A,\p^r J}^{(2)} = \sum_{s=0}^{2(r+1)} f_{r,s} \cos(2\pi s x).
\ee
On the other hand one can also write the $r\to+\inf$ conjecture (\ref{univRenyiCFT}) in Fourier series as
\be \label{fourier2}
1-2x+2x^2 = \f23 + \f{2}{\pi^2}\sum_{s=1}^{+\inf} \f{\cos(2\pi s x)}{s^2}.
\ee
We compare the Fourier coefficients in (\ref{fourier1}) and (\ref{fourier2}) in the right panel of Fig.~\ref{RenyiCFT}, and find good matches in the large $r$ limit.
It would be interesting to calculate the conjecture (\ref{univRenyiCFT}) explicitly in CFT.

\subsection{Harmonic chain}

The single-interval ground state R\'enyi entropy in the harmonic chains has been calculated in
\cite{Bombelli:1986rw,Srednicki:1993im,Callan:1994py,Peschel:1999DensityMatrices,Chung2000Densitymatrix,peschel2003calculation,%
Plenio:2004he,Cramer:2005mx,Casini:2005zv,Casini:2009sr,peschel2009reduced,peschel2012special} and one can calculate excited state R\'enyi entropy using the wave function method discussed further in \cite{Castro-Alvaredo:2018dja,Castro-Alvaredo:2018bij}.
In this subsection we elaborate on how to calculate the second R\'enyi entropy in the ground and excited state using the mini version of the wave function method which will lead to relatively  compact formulas.

We choose the subsystem $A=[1,\ell]$ and its complement $B=[\ell+1,L]$, and decompose the coordinates $Q=(Q_A,Q_B)$ with $Q_A=(q_1,\cdots,q_\ell)$ and $Q_B=(q_{\ell+1},\cdots,q_L)$ respectively.
Correspondingly, we decompose the matrices $W$ and $V_k$ defined in Section~\ref{secHC} as
\be
W = \Big( \ba{cc} A & B \\ C & D \ea \Big), ~~
V_k = \Big( \ba{cc} E_k & F_k \\ G_k & H_k \ea \Big).
\ee
The matrix $W$ is real symmetric, consequently the matrices $A$ and $D$ are also real symmetric, $B$ and $C$ are real, and $B^T=C$.
The matrix $V_k$ is hermitian, and as a result $E_k$ and $H_k$ are hermitian and $G_k^\dag=F_k$.

By integrating out the degrees of freedom of $B$, i.e. the coordinates $Q_B$, we get the RDM of the subsystem $A$
\be
\lag Q_A | \r_A | Q_A' \rag = \int \rD Q_B \lag Q_A,Q_B | \r | Q_A',Q_B\rag.
\ee
For the ground state density matrix (\ref{rhoG}) we get the RDM \cite{Bombelli:1986rw}
\be \label{rhoAG}
\lag Q_A | \r_{A,G} | Q_A' \rag = \sr{\det\f{\td A}{\pi}} \ep^{-\f12Q_A^TAQ_A-\f12Q_A'^TAQ_A'+\f14(Q_A+Q_A')^TBD^{-1}C(Q_A+Q_A')},
\ee
where we have defined
\be \label{tdA}
\td A = A - BD^{-1}C.
\ee
For the excited state density matrix (\ref{rhok}) we get
\bea \label{rhoAk}
&& \lag Q_A | \r_{A,k} | Q_A' \rag = \lag Q_A | \r_{A,G}| Q_A' \rag %\sr{\det\f{\td A}{\pi}} \ep^{-\f12x^TAx-\f12x'^TAx'+\f14(x+x')^TBD^{-1}C(x+x')}
                              \Big[
                                    \f12 \tr(H_kD^{-1})
                                  + Q_A^T E_k Q_A'
                                  - \f12 (Q_A+Q_A')^T BD^{-1}G_k Q_A' \\
&& \phantom{\lag Q_A | \r_{A,k} | Q_A' \rag =}
                                  - \f12 Q_A^T F_k D^{-1} C (Q_A+Q_A')
                                  + \f14 (Q_A+Q_A')^T BD^{-1}H_kD^{-1}C(Q_A+Q_A')
                                    \Big]. \nn
\eea
Following \cite{Peschel:1999DensityMatrices,peschel2012special}, one can also writes the RDMs in the operator form as follows:
\be \label{rhoAGoperators}
\r_{A,G} = 2^\ell \sr{ \det[ A (BD^{-1}C)^{-1} - 1 ] } \ep^{-\f12Q_A^T\td AQ_A} \ep^{-P_A^T (BD^{-1}C)^{-1} P_A} \ep^{-\f12Q_A^T\td AQ_A},
\ee
\bea
\label{rhoAkoperators}
&&
\r_{A,k} = \f12 \tr(H_kD^{-1}) \r_{A,G}
            - \f12 Q_A^T \hat F_k D^{-1} C Q_A \r_{A,G} %\nn\\
%&& \phantom{\r_{A,k} =}
            - \f12 \r_{A,G} Q_A^T BD^{-1} \hat G_k Q_A \nn\\
&& \phantom{\r_{A,k} =}
            + \sum_{j_1,j_2=1}^\ell [\hat E_k]_{j_1j_2} [Q_A]_{j_1} \r_{A,G} [Q_A]_{j_2},
\eea
where we have defined the matrices
\bea
&& \hat F_k = F_k-\f12BD^{-1}H_k, ~~
   \hat G_k = G_k-\f12H_k D^{-1} C, \nn\\
&& \hat E_k = E_k - \f12 BD^{-1}G_k - \f12 F_kD^{-1}C + \f12 BD^{-1}H_kD^{-1}C.
\eea
Note that $Q_A$, $P_A$ in (\ref{rhoAGoperators}) and (\ref{rhoAkoperators}) are understood as operators $\hat Q_A=(\hat q_1,\cdots,\hat q_\ell)$, $\hat P_A=(\hat p_1,\cdots,\hat p_\ell)$, and the orders of the terms are important.

To calculate the second R\'enyi entropy, we need to calculate the moments of the RDM
\be \label{trArhoA2}
\tr_A\r_A^2 = \int \rD Q_A \rD Q_A' \lag Q_A | \r_A | Q_A' \rag \lag Q_A' | \r_A | Q_A \rag.
\ee
For the ground state RDM (\ref{rhoAG}) we get
\be \label{trArAG2}
\tr_A \r_{A,G}^2 = \sr{\f{\det\td A}{ \det A}},
\ee
and for the excited state RDM (\ref{rhoAk}) we get
\bea \label{trArAk2}
&& \f{\tr_A \r_{A,k}^2}{\tr_A \r_{A,G}^2} =
                  \f18 \tr[ (\Re\td E_k)\td A^{-1}(\Re\td E_k)\td A^{-1} ]
                + \f18 \tr[ (\Re E_k) A^{-1}(\Re E_k) A^{-1} ] \nn\\
&& \phantom{\f{\tr_A \r_{A,k}^2}{\tr_A \r_{A,G}^2} =}
                - \f14 \tr[ \Im( \td E_k' )A^{-1}\Im(\td E_k'' )\td A^{-1} ]
                + X_k^2,
\eea
where we have defined
\bea \label{tdEk}
&& \td E_k = E_k - BD^{-1}G_k - F_kD^{-1}C + BD^{-1}H_kD^{-1}C, \nn\\
&& \td E_k' = E_k-BD^{-1}G_k , ~~
   \td E_k'' = E_k-F_kD^{-1}C, \nn\\
&& X_k = \f14 \tr(\td E_k \td A^{-1}) - \f14 \tr(E_k A^{-1}) + \f12 \tr(H_k D^{-1}).
\eea

According to \cite{Castro-Alvaredo:2018dja,Castro-Alvaredo:2018bij,Castro-Alvaredo:2019irt,Castro-Alvaredo:2019lmj} there must be a universal form in the large momentum limit
\be \label{CDDS}
\lim_{|k| \to +\inf} \cF_{A,k}^{(2)} = x^2 + (1-x)^2,
\ee
which should be valid even for a very small gap $m$. In CFT, it is just (\ref{univRenyiCFT}).
Note that for the universal R\'enyi entropy to be valid, one also needs to impose the continuum limit $L\to+\inf$ and $\ell\to+\inf$ with fixed $x=\f{\ell}{L}$.
This has been checked extensively in \cite{Castro-Alvaredo:2018dja,Castro-Alvaredo:2018bij}, and we will not repeat it here.

\begin{figure}[tp]
  \centering
  % Requires \usepackage{graphicx}
  \includegraphics[height=0.24\textwidth]{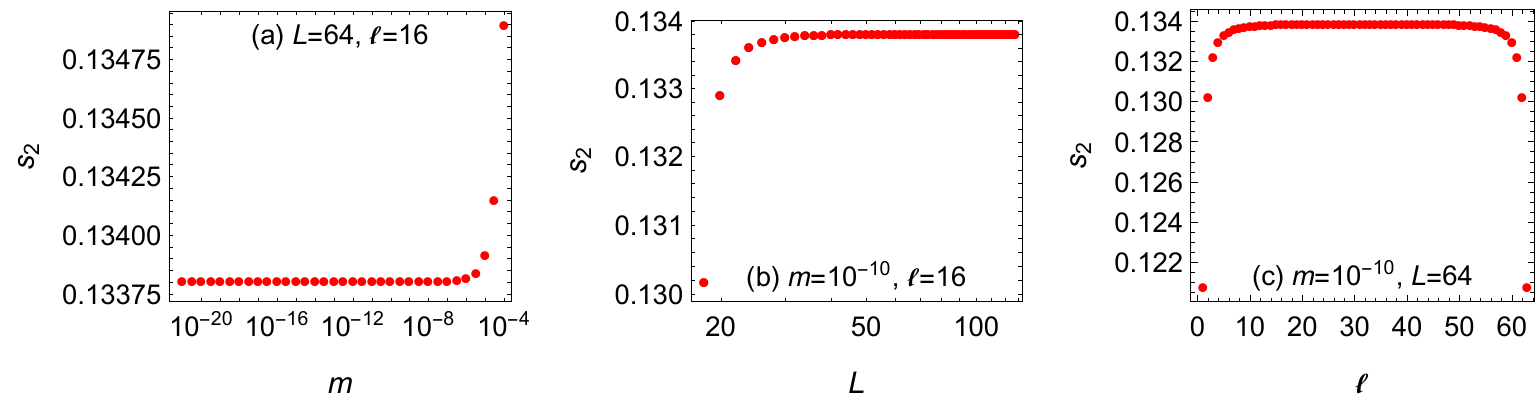}\\
  \caption{The constant term of the single-interval ground state second R\'enyi entropy (\ref{SAG2}) is independent of $m$, $L$, $\ell$ in the massless and continuum limit of the harmonic chain. We read the approximate value $s_2 \approx 0.134$.}\label{RenyiGS}
\end{figure}

\begin{figure}[p]
  \centering
  % Requires \usepackage{graphicx}
  \includegraphics[height=0.9\textwidth]{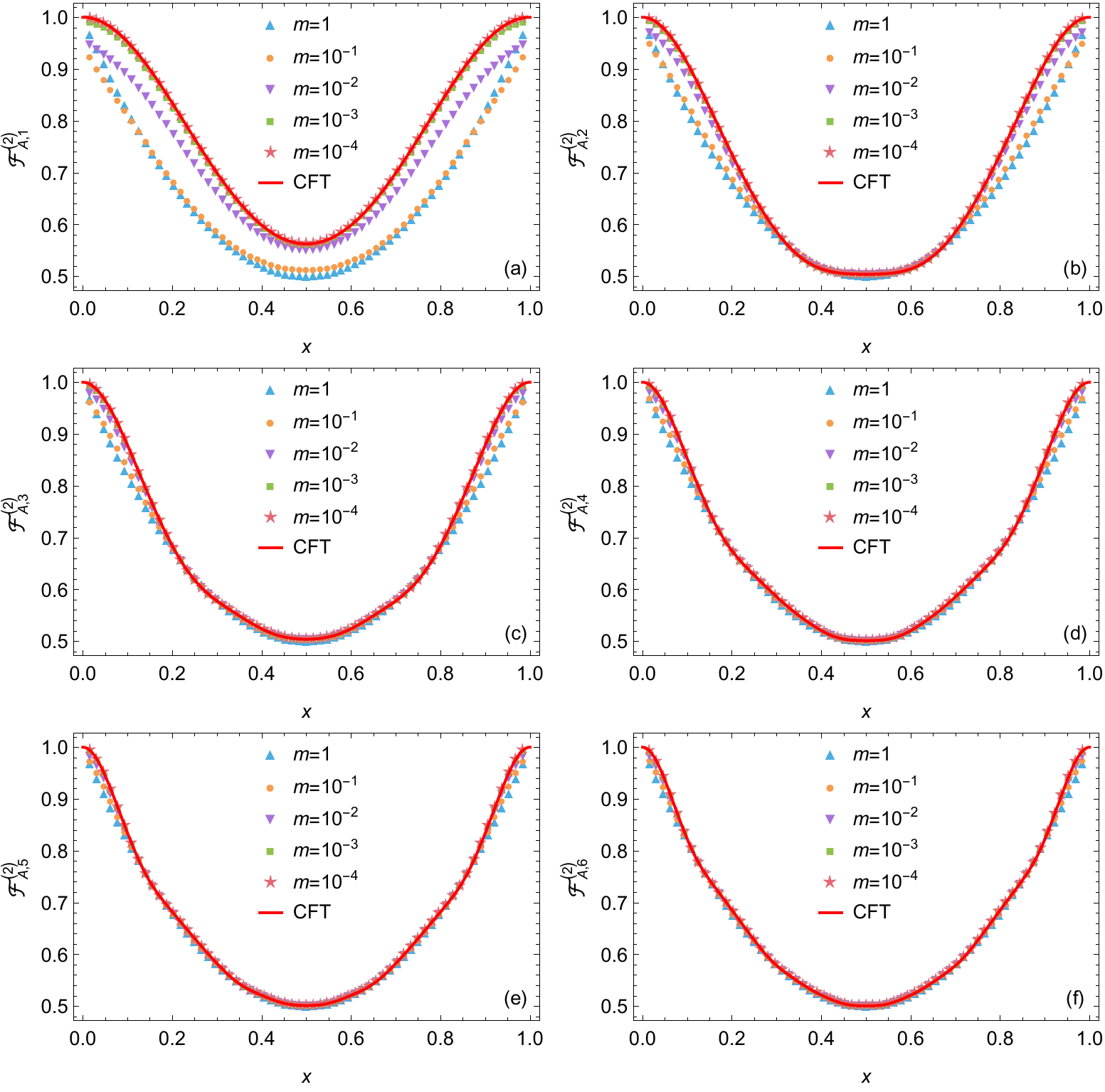}\\
  \caption{The excited state single-interval R\'enyi entropies of the harmonic chain (symbols) and the 2D free massless bosonic theory (lines). There are perfect matches in the gapless limit $m\to0$ (\ref{matchRenyi}). We have set $L=64$.}\label{Renyik}
\end{figure}

We also calculate the constant term $s_2$ in (\ref{SAG2}) in Fig.~\ref{RenyiGS}. We read easily in the figure that the constant $s_2 \approx 0.134$.
We compare the single-particle excited state R\'enyi entropies in lattice and CFT in Fig.~\ref{Renyik}.
We see that there are perfect matches of the lattice and CFT results in the massless limit
\be \label{matchRenyi}
\lim_{m\to0} \cF_{A,k}^{(2)} = \cF_{A,\p^{k-1} J} ^{(2)}, ~ k=1,2,\cdots.
\ee
As expected, in each of the excited state R\'enyi entropies in the harmonic chain there is the same IR divergent term $\f12\log\f{1}{mL}$, i.e. that in the ground state R\'enyi entropy (\ref{SAG2}).

\section{Schatten distance}\label{secSchatten}

We consider the Schatten distance between the RDMs of the excited states of the current and its derivatives in the 2D free massless non-compact bosonic theory.
We also calculate the Schatten distance between the single-interval RDMs of the ground state and the single-particle states in the gapless limit of the harmonic chain.
We find a universal from of the distance in the limit of both large momenta and large momentum difference and a new corrected from of the distance when there is only the limit of large momentum difference.

\subsection{Massless bosonic theory}

To calculate the second Schatten distance in 2D free massless bosonic theory, besides (\ref{fourpointfunction1}), we need the four-point functions on the two-fold plane
\bea
&& \f{\tr_A( \r_{A,\p^r J} \r_{A,\p^s J} )}{\tr_A\r_{A,G}^2} = \f{1}{r!s!(r+1)!(s+1)!} \lag {\p^rJ(0_1)} {\p^rJ(\inf_1)} {\p^sJ(0_2)} {\p^sJ(\inf_2)} \rag_{\rC^2}, \nn\\
&& \f{\tr_A( \r_{A,\bar\p^r \bar J} \r_{A,\bar\p^s \bar J} )}{\tr_A\r_{A,G}^2} =
      \f{1}{r!s!(r+1)!(s+1)!} \lag {\bar\p^r\bar J(0_1)} {\bar\p^r\bar J(\inf_1)}
                                   {\bar \p^s\bar J(0_2)} {\bar\p^s\bar J(\inf_2)} \rag_{\rC^2}, \nn\\
&& \f{\tr_A( \r_{A,\p^r J} \r_{A,\bar\p^s \bar J} )}{\tr_A\r_{A,G}^2} =
      \f{1}{r!s!(r+1)!(s+1)!} \lag {\p^rJ(0_1)} {\p^rJ(\inf_1)}
                                   {\bar \p^s\bar J(0_2)} {\bar\p^s\bar J(\inf_2)} \rag_{\rC^2}.
\eea
Note that we have the factorization
\be
\lag {\p^rJ(0_1)} {\p^rJ(\inf_1)} {\bar \p^s\bar J(0_2)} {\bar\p^s\bar J(\inf_2)} \rag_{\rC^2}
= \lag {\p^rJ(0_1)} {\p^rJ(\inf_1)} \rag_{\rC^2}
  \lag {\bar \p^s\bar J(0_2)} {\bar\p^s\bar J(\inf_2)} \rag_{\rC^2}.
\ee
We finally get
\be
D_2(\r_{A,G},\r_{A,J})^2 = \f{1}{256} [ 99 - 128 \cos(\pi x) + 28 \cos(2 \pi x) + \cos(4 \pi x) ],
\ee
which has been calculated in \cite{Zhang:2019itb}, as well as the new results that we collect in Appendix~\ref{appSchatten}.
It is easy to see in CFT we have
\bea
&& D_2(\r_{A,G},\r_{A,\p^r J})=D_2(\r_{A,G},\r_{A,\bar\p^r \bar J}), \nn\\
&& D_2(\r_{A,\p^r J},\r_{A,\p^s J})=D_2(\r_{A,\bar\p^r \bar J},\r_{A,\bar\p^s \bar J}),\nn\\
&& D_2(\r_{A,\p^r J},\r_{A,\bar\p^s \bar J})=D_2(\r_{A,\p^s J},\r_{A,\bar\p^r \bar J}),
\eea
just like on the lattice there are
\bea
&& D_2(\r_{A,G},\r_{A,k})=D_2(\r_{A,G},\r_{A,-k}), \nn\\
&& D_2(\r_{A,k},\r_{A,l})=D_2(\r_{A,-k},\r_{A,-l}).
\eea

\subsection{Harmonic chain}

To calculate the second Schatten distance in the harmonic chain, except the momentum (\ref{trArhoA2}), we also need the product
\be
\tr_A(\r_A\s_A) = \int \rD Q_A \rD Q_A' \lag Q_A | \r_A | Q_A' \rag \lag Q_A' | \s_A | Q_A \rag.
\ee
After straightforward but lengthy calculation, we obtain
\be
\f{\tr_A(\r_{A,G}\r_{A,k})}{\tr_A \r_{A,G}^2} = X_k,
\ee
as well as
\bea
&& \f{\tr_A(\r_{A,k_1}\r_{A,k_2})}{\tr_A \r_{A,G}^2} =
         \f18 \tr[ (\Re\td E_{k_1})\td A^{-1}(\Re\td E_{k_2})\td A^{-1} ]
       + \f18 \tr[ (\Re E_{k_1}) A^{-1}(\Re E_{k_2}) A^{-1} ] \nn\\
&& \phantom{\f{\tr_A(\r_{A,k_1}\r_{A,k_2})}{\tr_A \r_{A,G}^2}=}
       - \f14 \tr[ \Im( \td E_{k_1}' )A^{-1}\Im( \td E_{k_2}'' )\td A^{-1} ]
       +X_{k_1}X_{k_2},
\eea
with the definitions (\ref{tdA}) and (\ref{tdEk}).

\begin{figure}[p]
  \centering
  % Requires \usepackage{graphicx}
  \includegraphics[height=0.9\textwidth]{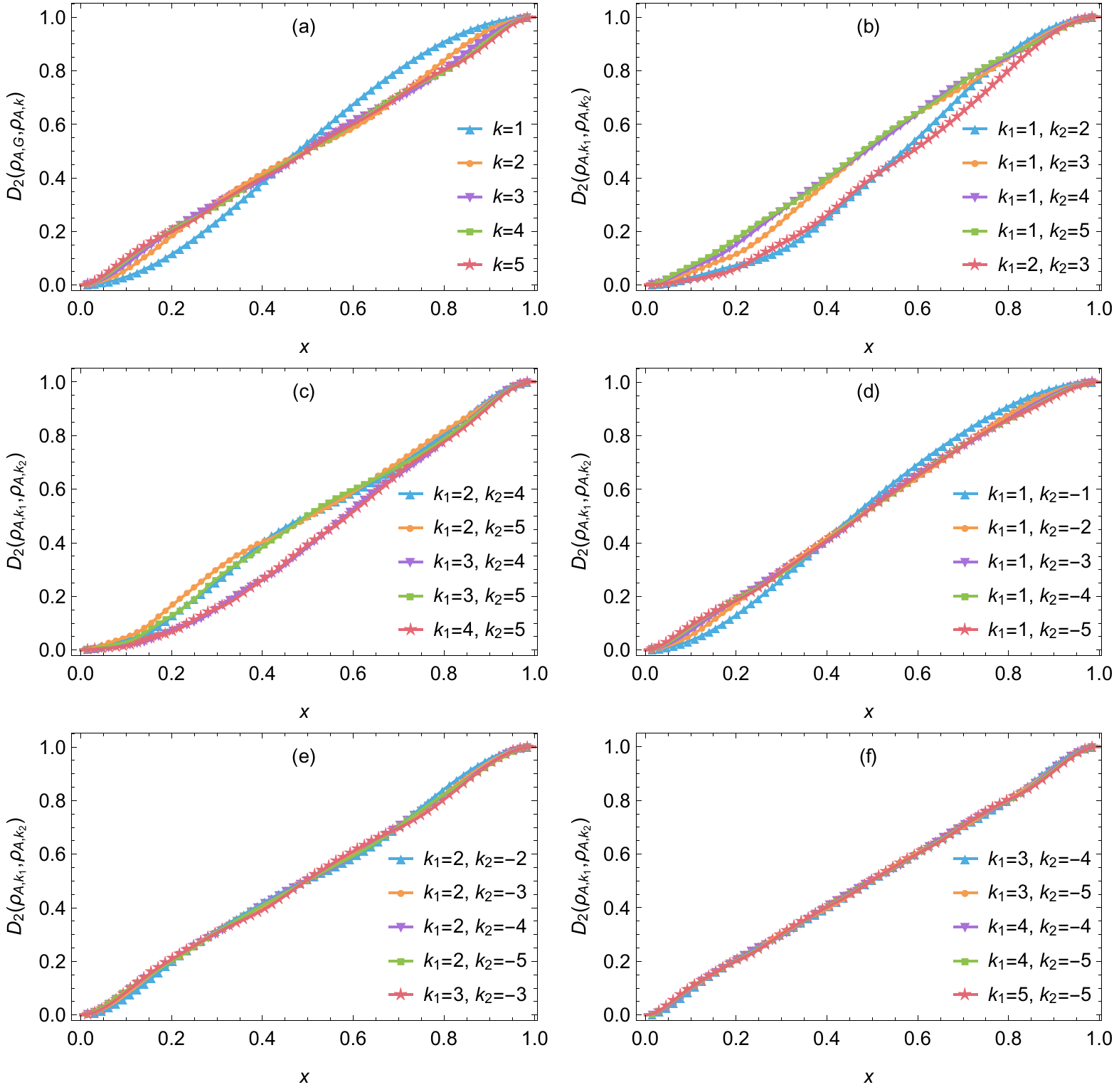}\\
  \caption{The second Schatten distances between the ground and excited states in the massless limit of the harmonic chain (symbols) and the 2D free massless bosonic theory (lines). There are perfect matches of the numerical lattice and analytical CFT results (\ref{mathchSchatten}). We have set $m=10^{-5}$, $L=64$.}\label{Schatten}
\end{figure}

We compare the lattice and CFT results in Fig.~\ref{Schatten} and find perfect matches
\bea \label{mathchSchatten}
&& \lim_{m \to 0} D_2( \r_{A,G}, \r_{A,k} ) = D_2( \r_{A,G}, \r_{A,\p^{k-1}J} ), \nn\\
&& \lim_{m \to 0} D_2( \r_{A,k_1}, \r_{A,k_2} ) = D_2( \r_{A,\p^{k_1-1}J}, \r_{A,\p^{k_2-1}J} ), \nn\\
&& \lim_{m \to 0} D_2( \r_{A,k_1}, \r_{A,-k_2} ) = D_2( \r_{A,\p^{k_1-1}J}, \r_{A,\bar\p^{k_2-1}\bar J} ),
\eea
where $k,k_1,k_2=1,2,\cdots$.

\begin{figure}[t]
  \centering
  % Requires \usepackage{graphicx}
  \includegraphics[height=0.6\textwidth]{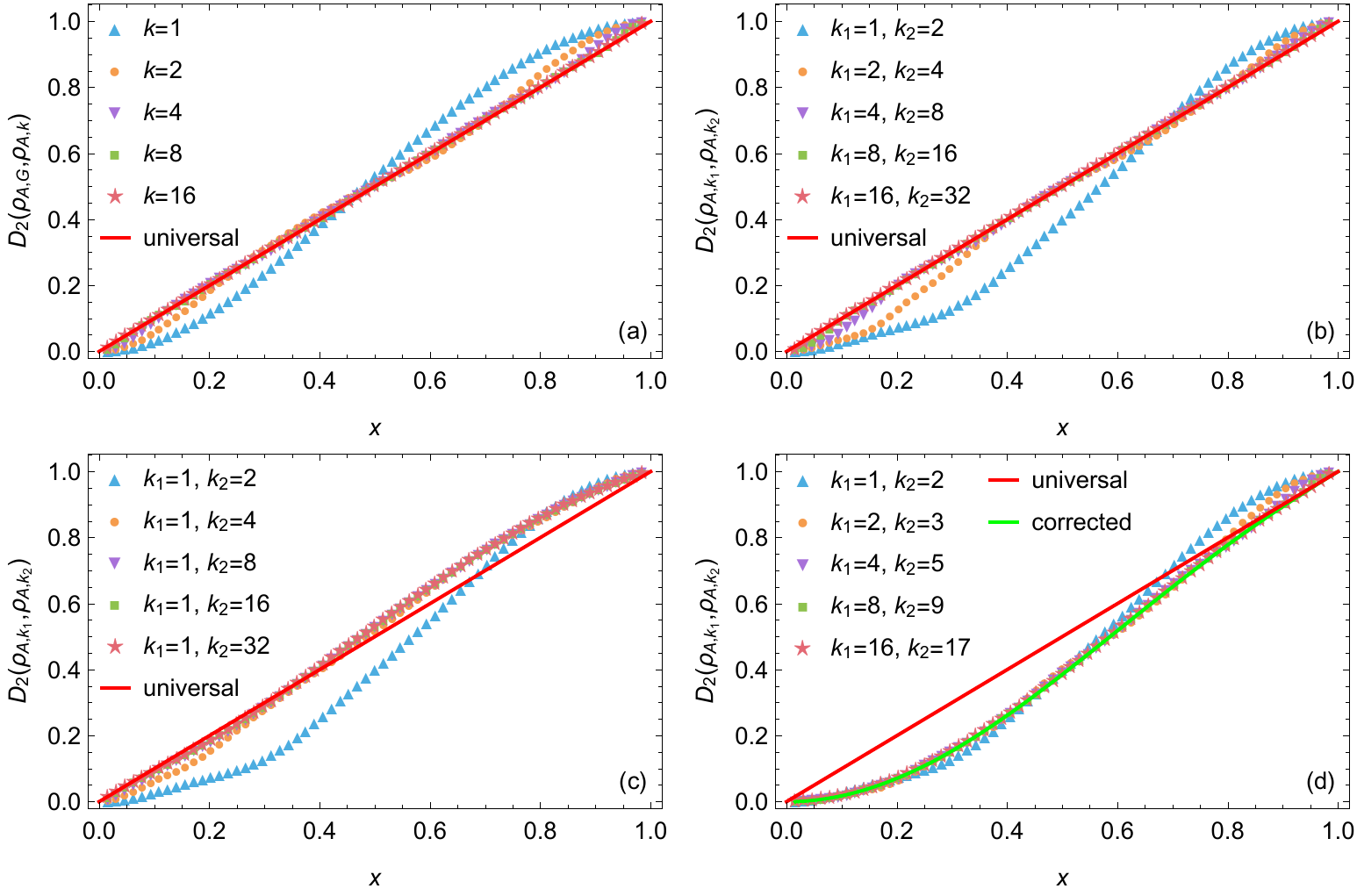}\\
  \caption{The numerical results of the second Schatten distance between the ground state and the single-particle states in the massless limit of the harmonic chain (symbols).
  For comparison, we also plot the universal distance $D_2^{\rm univ}$ (\ref{D2univ}) (red lines) and the corrected distance $D_{2,k}^{\rm corr}$ (\ref{D2corr}) (green line). We have set $m=10^{-5}$, $L=64$.}
  \label{SchattenUnivCorr}
\end{figure}

We show the numerical results of the second Schatten distance for states with large momenta in Fig.~\ref{SchattenUnivCorr}.
We find the asymptotic universal behavior
\bea \label{Duniv}
&& \lim_{|k| \to +\inf} D_2( \r_{A,G}, \r_{A,k}) = D_2^{\rm univ}, \label{DunivGk} \\
&& \lim_{|k_1|\to+\inf,|k_2|\to+\inf,|k_1-k_2|\to+\inf} D_2( \r_{A,k_1}, \r_{A,k_2}) = D_2^{\rm univ}, \label{Dunivk1k2} \\
&& \lim_{|k_1|\to+\inf,|k_2|\to+\inf} D_2( \r_{A,k_1}, \r_{A,k_2}) = D_{2,k_1-k_2}^{\rm corr},
\eea
with the universal distance and the new corrected form
\be \label{D2univ}
D_2^{\rm univ} = x,
\ee
\be \label{D2corr}
D_{2,k}^{\rm corr} = x \sr{ 1 - \f{\sin^2({\pi k\ell}/{L})}{\ell^2\sin^2({\pi k}/{L})} }.
\ee
For $D_2( \r_{A,G}, \r_{A,k})$ to take the universal form $D_2^{\rm univ}$, it is enough to consider the large momentum limit $|k|\to+\inf$.
For $D_2( \r_{A,k_1}, \r_{A,k_2})=D_2^{\rm univ}$, we need not only the large momentum limit $|k_1|\to+\inf$ and $|k_2|\to+\inf$, but also the limit of large momentum difference $|k_1-k_2|\to+\inf$.
For $D_2( \r_{A,k_1}, \r_{A,k_2})=D_{2,k_1-k_2}^{\rm corr}$, we need only the large momentum limit $|k_1|\to+\inf$ and $|k_2|\to+\inf$.

The universal Schatten distance $D_2^{\rm univ}$ (\ref{D2univ}) could be explained in terms of quasiparticles, similar to the universal R\'enyi entropy in \cite{Castro-Alvaredo:2018dja,Castro-Alvaredo:2018bij,Castro-Alvaredo:2019irt,Castro-Alvaredo:2019lmj}, as we have reviewed around (\ref{rAkx111mx00}).
We use $|00]$ to denote the state of the subsystem $A$ with no quasiparticle, $|10]$ to denote the state of $A$ with the quasiparticle with momentum $k_1$, and $|01]$ to denote the state with the quasiparticle with momentum $k_2$. Then there are the RDMs
\be
\r_{A,G} = |00][00], ~~
\r_{A,k_1} = x |10][10| + (1-x) |00][00], ~~
\r_{A,k_2} = x |01][01| + (1-x) |00][00],
\ee
which give easily (\ref{DunivGk}) and (\ref{Dunivk1k2}).

The corrected result of the Schatten distance $D_2^{\rm corr}$ is derived in the extremely gapped limit by writing the excited state in terms of local excitations, but it is still valid in the slightly gapped harmonic chain in the large momentum limit, just like the corrections to the universal R\'enyi entropy in \cite{Zhang:2020vtc,Zhang:2020dtd}.
We will report the derivations of the corrected result (\ref{D2corr}) in \cite{ZRDistance}.

\section{R\'enyi mutual information}\label{secRMI}

In this section we consider the R\'enyi mutual information of two disjoint intervals with lengths $\ell_1$ and $\ell_2$ and distance $d$ on a circle with length $L$ in the 2D free massless non-compact bosonic theory and the gapless limit of the harmonic chain.
It is convenient to define $x_1=\f{\ell_1}{L}$, $x_2=\f{\ell_2}{L}$, $y=\f{d}{L}$.
We verify the universal IR divergent term in the ground and excited state double-interval R\'enyi entropy.

\subsection{Massless bosonic theory}

For two intervals $A=A_1\cup A_2$ with $A_1=[0,\ell_1]$ and $A_2=[\ell_1+d,\ell_1+d+\ell_2]$, one can calculate the R\'enyi entropies $S_{A_1}^{(2)}$, $S_{A_2}^{(2)}$, $S_{A_1A_2}^{(2)}$ and define the R\'enyi mutual information
\be
I^{(2)}_{A_1A_2} = S_{A_1}^{(2)} + S_{A_2}^{(2)} - S_{A_1A_2}^{(2)}.
\ee
In the ground state of 2D free massless compact bosonic theory on a cylinder with circumference $L$, the second R\'enyi mutual information is \cite{Furukawa:2008uk,Calabrese:2009ez,Headrick:2010zt}
\be
I_{A_1A_2,G}^{(2)} = \f14 \log \f{\sin\f{\pi(\ell_1+d)}{L}\sin\f{\pi(\ell_2+d)}{L}}{\sin\f{\pi d}{L}\sin\f{\pi(\ell_1+d+\ell_2)}{L}}
               +\log \f{\th_3(\eta\t)\th_3(\t/\eta)}{[\th_3(\t)]^2},
\ee
where $\eta$ is related to the radius of compact boson target space $R$ as $\eta=\f{R^2}2$, the purely imaginary parameter $\t$ is determined by the cross ratio as
\be
\f{\sin\f{\pi \ell_1}{L}\sin\f{\pi \ell_2}{L}}{\sin\f{\pi (\ell_1+d)}{L}\sin\f{\pi(d+\ell_2)}{L}} = \Big[ \f{\th_2(\t)}{\th_3(\t)} \Big]^4,
\ee
and $\th_2(\t)$, $\th_3(\t)$ are the usual theta functions
\be
\th_2(\t) = \sum_{r\in\rZ}\ep^{\pi\ii\t(r+\f12)^2}, ~~
\th_3(\t) = \sum_{r\in\rZ}\ep^{\pi\ii\t r^2}.
\ee
In the non-compact limit $\eta\to+\inf$, the R\'enyi mutual information becomes
\be \label{IA1A2G2}
I_{A_1A_2,G}^{(2)} = \f14 \log \f{\sin\f{\pi(\ell_1+d)}{L}\sin\f{\pi(\ell_2+d)}{L}}{\sin\f{\pi d}{L}\sin\f{\pi(\ell_1+d+\ell_2)}{L}}
               +\log \f{1}{\sr{-\ii\t}[\th_3(\t)]^2}
               +\f12\log\eta.
\ee
On the RHS there are three terms; from left to right: the universal part, the specific part that depends on the state, and the IR divergent part.
The free massless non-compact bosonic theory can be also viewed as the massless limit of the massive theory, and the R\'enyi mutual information is expected to also have three parts with the universal and specific parts the same as those in (\ref{IA1A2G2}) and the IR divergent part dependent on the infinitesimal mass $m$.
In \cite{Casini:2009sr}, it was argued that the IR divergent term in the R\'enyi entropy is independent of the number of the intervals, and then we expect the same IR divergent term in the R\'enyi mutual information as in the R\'enyi entropy.
By guessing and considering the single interval R\'enyi entropy (\ref{SAG2}) we anticipate that
\be\label{IA1A2G2m}
I_{A_1A_2,G}^{(2)} = \f14 \log \f{\sin\f{\pi(\ell_1+d)}{L}\sin\f{\pi(\ell_2+d)}{L}}{\sin\f{\pi d}{L}\sin\f{\pi(\ell_1+d+\ell_2)}{L}}
               +\log \f{1}{\sr{-\ii\t}[\th_3(\t)]^2}
               +\f12\log\f{1}{m L}.
\ee
In a general state we define the subtracted mutual information
\be
J_{A_1A_2}^{(2)} = I_{A_1A_2}^{(2)}
               - \f14 \log \f{\sin\f{\pi(\ell_1+d)}{L}\sin\f{\pi(\ell_2+d)}{L}}{\sin\f{\pi d}{L}\sin\f{\pi(\ell_1+d+\ell_2)}{L}}
               - \f12\log\f{1}{m L},
\ee
which we anticipate is independent of the mass $m$ in the massless limit.
For the ground state, it is just
\be \label{JA1A2G2}
J_{A_1A_2,G}^{(2)} = \log \f{1}{\sr{-\ii\t}[\th_3(\t)]^2}.
\ee

\subsection{Harmonic chain}

The multi-interval R\'enyi entropy in the harmonic chain has been already considered for the ground state \cite{Coser:2013qda,DeNobili:2015dla} and the excited states \cite{Castro-Alvaredo:2019irt}.
The wave function method can be easily adapted to the multi-interval case by just relabeling the sites on the chain \cite{Castro-Alvaredo:2019irt}.
We choose two disjoint intervals $A=A_1\cup A_2$ with $A_1=[1,\ell_1]$ and $A_2=[\ell_1+d+1,\ell_1+d+\ell_2]$, and calculate the R\'enyi entropy of $A$ with its complement $B$.
Then we can use (\ref{trArAG2}) and (\ref{trArAk2}) to calculate the double-interval R\'enyi entropy $S_{A_1A_2}^{(2)}$ in the ground state and single-particle excited states, from which we get the R\'enyi mutual information $I^{(2)}_{A_1A_2}$ of $A_1$ and $A_2$.
According to \cite{Castro-Alvaredo:2018dja,Castro-Alvaredo:2018bij,Castro-Alvaredo:2019irt,Castro-Alvaredo:2019lmj} there is a universal form in the large momentum limit
\be
\lim_{|k| \to +\inf} \cF_{A_1A_2,k}^{(2)} = (x_1+x_2)^2 + (1-x_1-x_2)^2.
\ee
This has been checked in \cite{Castro-Alvaredo:2019irt}, and we will not repeat it here.

\begin{figure}[htbp]
  \centering
  % Requires \usepackage{graphicx}
  \includegraphics[height=0.6\textwidth]{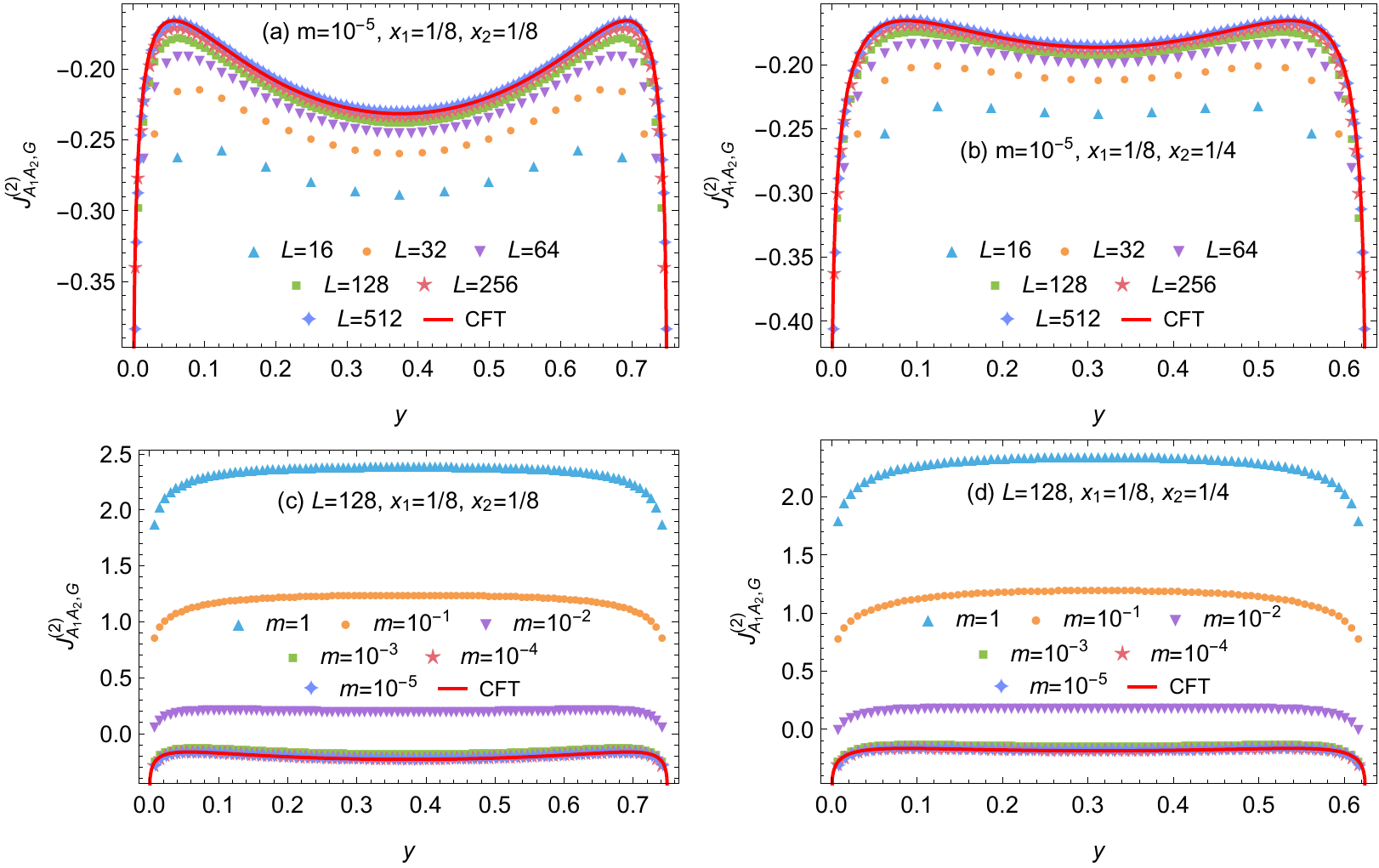}\\
  \caption{The second R\'enyi mutual information  in ground state of the harmonic chain in the continuum limit and the massless limit. We verify (\ref{JA1A2G2}). Especially, we verify the IR divergent term in the ground state double-interval R\'enyi entropy (\ref{IA1A2G2m}).}
  \label{mutualG}
\end{figure}

\begin{figure}[htbp]
  \centering
  % Requires \usepackage{graphicx}
  \includegraphics[height=0.3\textwidth]{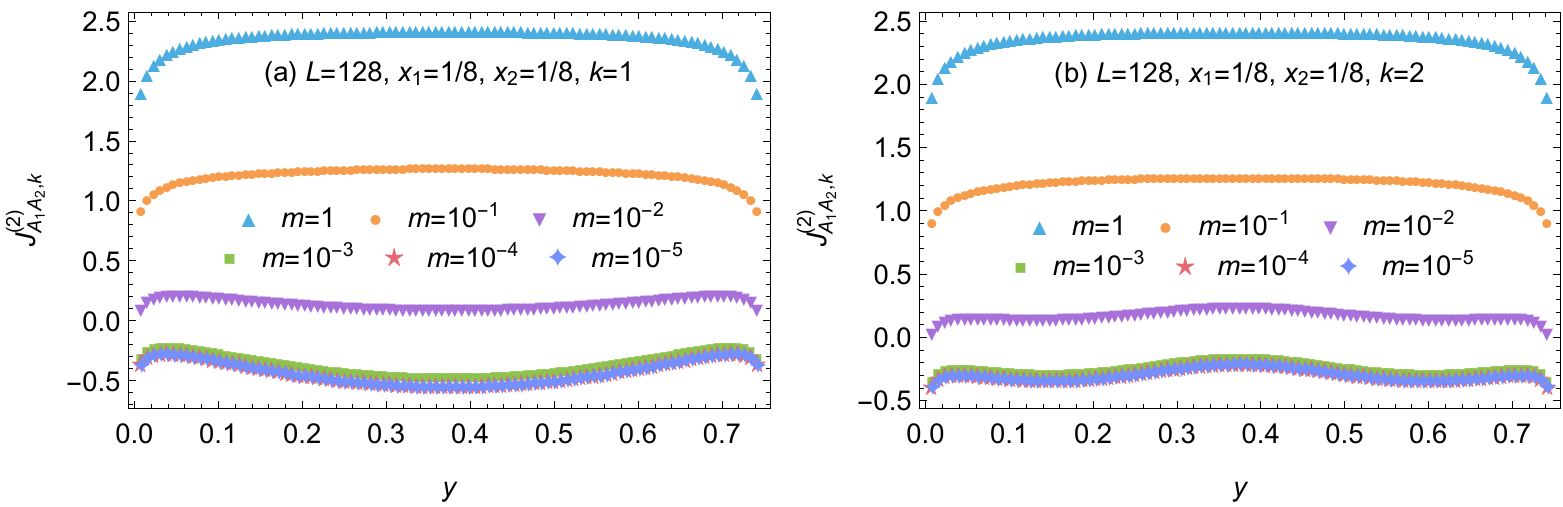}\\
  \caption{The second R\'enyi mutual information in the single-particle state of massless limit of the harmonic chain. We verify that the excited state double-interval R\'enyi entropy has the same IR divergent term as that in the ground state.}
  \label{mutualk}
\end{figure}

We check the CFT prediction (\ref{JA1A2G2}) in Fig.~\ref{mutualG}, where we plot the numerical lattice R\'enyi mutual information and the analytical result in the continuum limit and the massless limit.
In Fig.~\ref{mutualk}, we plot the subtracted mutual information in the single-particle excited state in the massless limit of the harmonic chain.
We see that it approaches to a fixed finite result in the massless limit.

\section{Conclusion}\label{secCon}

We have calculated analytically the R\'enyi entropy in the excited states of the current and its derivatives in the 2D free massless non-compact bosonic theory and the subsystem Schatten distance between these excited states.
We also calculated numerically the same quantities for the single-particle excited states of the short-range coupled harmonic oscillators in the gapless limit.
The lattice numerical results coming from the excited states of the harmonic chain match perfectly with the analytical CFT results of the bosonic theory.
We have focused on the second R\'enyi entropy and the second Schatten distance in the single-particle excited states. To calculate the same quantities efficiently for multi-particle excited states of the harmonic chains one needs to use the full-fledged wave function method as in \cite{Castro-Alvaredo:2018dja,Castro-Alvaredo:2018bij}, and in the corresponding CFT one needs to consider higher level descendant states.
We will come back to this problem in an upcoming work \cite{Zhang:2020txb}.

In the limit of both large momenta and large momentum difference, we found a universal Schatten distance (\ref{D2univ}) that is independent of the momenta.
However, when we consider only the limit of large momenta but with a small momentum difference, one reaches to a more complicated result (\ref{D2corr}).
We will elaborate the derivations of the universal and corrected Schatten distances elsewhere \cite{ZRDistance}.
In the limit of large momenta, one expects to be a universal R\'enyi entropy for the quasiparticle excited states
\cite{Castro-Alvaredo:2018dja,Castro-Alvaredo:2018bij,Castro-Alvaredo:2019irt,Castro-Alvaredo:2019lmj}, and we have verified this for the single-particle excited states in the massless limit of the harmonic chain.
However, in the multi-particle excited states there exist corrections to the universal R\'enyi entropy when the momentum differences are small, as we report in \cite{Zhang:2020vtc,Zhang:2020dtd}.

In this paper we have made some preliminary investigations of the excited state R\'enyi mutual information in the massless noncompact bosonic theory.
We could only show numerically in the harmonic chain that the subtracted excited state R\'enyi mutual information approaches to a fixed finite result in the massless and continuum limit, but we have not derived an analytical expression in the field theory.
It is also interesting to investigate the excited state R\'enyi mutual information when the boson is compact.
The 2D free massless compact bosonic theory has the discrete realization of the spin-$\f12$ XXZ chain or the Ashkin-Teller chain, depending on whether it is simple compactified or orbifold compactified.
Still, the analytical calculation of the excited state R\'enyi mutual information in the 2D free massless compact bosonic theory is difficult.
To calculate the excited state R\'enyi mutual information with index $n$ in a general 2D CFT, one needs to evaluate a six-point correlation function in the $n$-fold CFT, with four of the operators being the twist operators and two of the operators being the one that generates the excited states, or equivalently one evaluates the $2n$-point correlation function of operator that generates the excited states in the CFT on a genus $n-1$ Riemann surface.
We hope to come back to the problem in the future.

\section*{Acknowledgements}

We thank Pasquale Calabrese for reading an early version of the draft and helpful discussions, comments and suggestions.
We also thank Sara Murciano for helpful discussions.
MAR thanks CNPq and FAPERJ (grant number 210.354/2018) for partial support.
JZ acknowledges support from ERC under Consolidator grant number 771536 (NEMO).

\appendix

\section{Results of R\'enyi entropy in CFT}\label{appRenyi}

In this appendix, we collect the results of the R\'enyi entropy of the excited states of the current and its derivatives in the 2D free massless bosonic theory that are omitted in Section~\ref{secRenyi}.
We obtain the results the results $\cF_{A,\p^r J}^{(2)}$ with $r=0,1,\cdots,13$.
We only show $\cF_{A,\p^r J}^{(2)}$ with $r=0,1,\cdots,8$ as follows:
\be
\cF_{A,J}^{(2)} = \frac{1}{128} [ 99 + 28 \cos (2 \pi  x)+\cos (4 \pi  x) ],
\ee
\be
\cF_{A,\p J}^{(2)} = \frac{1}{32768}[22931+8072 \cos (2 \pi  x)+1628 \cos (4 \pi  x)+56 \cos (6 \pi  x)+81 \cos (8 \pi  x)],
\ee
\bea
&& \cF_{A,\p^2 J}^{(2)} = \frac{1}{524288}
      [ 358254
      +119016 \cos (2 \pi  x)
      +34431 \cos (4 \pi  x)
      +11044 \cos (6 \pi  x) \nn\\
&& \phantom{\cF_{A,\p^2 J}^{(2)} =}
      +930 \cos (8 \pi  x)
      -12 \cos (10 \pi  x)+625 \cos (12 \pi  x) ],%\nn%\\
\eea
\bea
&& \cF_{A,\p^3 J}^{(2)} = \frac{1}{2147483648}
      [1453496467+468409168 \cos (2 \pi  x)
      +131718088 \cos (4 \pi  x) \nn\\
&& \phantom{\cF_{A,\p^3 J}^{(2)} =}
       +64487152 \cos (6 \pi  x)
       +24720860 \cos (8 \pi  x)
       +2817488 \cos (10 \pi  x)
       +503800 \cos (12 \pi  x) \nn\\
&& \phantom{\cF_{A,\p^3 J}^{(2)} =}
       -170000 \cos (14 \pi  x)
       +1500625 \cos (16 \pi  x)],%\nn%\\
\eea
\bea
&& \cF_{A,\p^4 J}^{(2)} = \f{1}{34359738368}
[ 23144240154 + 7332632360 \cos(2 \pi x) +
  2007130130 \cos(4 \pi x) \nn\\
&& \phantom{\cF_{A,\p^4 J}^{(2)} =} + 980954800 \cos(6 \pi x)+
  589623400 \cos(8 \pi x)+ 248171248 \cos(10 \pi x)\nn\\
&& \phantom{\cF_{A,\p^4 J}^{(2)} =}+
  33050605 \cos(12 \pi x)+ 8911340 \cos(14 \pi x)+
  1240190 \cos(16 \pi x)\nn\\
&& \phantom{\cF_{A,\p^4 J}^{(2)} =}- 1968820 \cos(18 \pi x)+
  15752961 \cos(20 \pi x) ],
\eea
\bea
&& \cF_{A,\p^5 J}^{(2)} = \frac{1}{8796093022208}
      [5908410214094
      +1852490627568 \cos (2 \pi  x)
      +497350812456 \cos (4 \pi  x)\nn\\
&& \phantom{\cF_{A,\p^5 J}^{(2)} =}
      +239692699664 \cos (6 \pi  x)
      +145554461967 \cos (8 \pi  x)
      +97574091720 \cos (10 \pi  x)\nn\\
&& \phantom{\cF_{A,\p^5 J}^{(2)} =}
      +43480385732 \cos (12 \pi  x)
      +6361228344 \cos (14 \pi  x)
      +2064791106 \cos (16 \pi  x)\nn\\
&& \phantom{\cF_{A,\p^5 J}^{(2)} =}
      +623898968 \cos (18 \pi  x)
      -16855020 \cos (20 \pi  x)
      -340730712 \cos (22 \pi  x)\nn\\
&& \phantom{\cF_{A,\p^5 J}^{(2)} =}
      +2847396321 \cos (24 \pi  x)],%\nn\\
\eea
\bea
&& \cF_{A,\p^6 J}^{(2)} = \f{1}{140737488355328}
      [ 94367240743036
      +29385073390736 \cos (2 \pi  x)\nn\\
&& \phantom{\cF_{A,\p^6 J}^{(2)} =}
      +7780670276267 \cos (4 \pi  x)
      +3702341089468 \cos (6 \pi  x)
      +2233254151814 \cos (8 \pi  x)\nn\\
&& \phantom{\cF_{A,\p^6 J}^{(2)} =}
      +1521933418796 \cos (10 \pi  x)
      +1091626255769 \cos (12 \pi  x)
      +505399189848 \cos (14 \pi  x)\nn\\
&& \phantom{\cF_{A,\p^6 J}^{(2)} =}
      +78677347812 \cos (16 \pi  x)
      +28478160536 \cos (18 \pi  x)
      +11038480299 \cos (20 \pi  x)\nn\\
&& \phantom{\cF_{A,\p^6 J}^{(2)} =}
      +2956053492 \cos (22 \pi  x)
      -1319038182 \cos (24 \pi  x)
      -3752254044 \cos (26 \pi  x)\nn\\
&& \phantom{\cF_{A,\p^6 J}^{(2)} =}
      +33871089681 \cos (28 \pi  x) ],%\nn
\eea
\bea
&& \cF_{A,\p^7 J}^{(2)} = \f{1}{9223372036854775808}
      [6177040104007000211
      +1914335191313784736 \cos (2 \pi  x) \nn\\
&& \phantom{\cF_{A,\p^7 J}^{(2)} =}
      +501762775231181392 \cos (4 \pi  x)
      +236301107644828128 \cos (6 \pi  x) \nn\\
&& \phantom{\cF_{A,\p^7 J}^{(2)} =}
      +141375547787711944 \cos (8 \pi  x)
      +96111229204173856 \cos (10 \pi  x) \nn\\
&& \phantom{\cF_{A,\p^7 J}^{(2)} =}
      +70336647651896304 \cos (12 \pi  x)
      +52823685514601568 \cos (14 \pi  x) \nn\\
&& \phantom{\cF_{A,\p^7 J}^{(2)} =}
      +25132319926583772 \cos (16 \pi  x)
      +4086561300970016 \cos (18 \pi  x) \nn\\
&& \phantom{\cF_{A,\p^7 J}^{(2)} =}
      +1590279016776912 \cos (20 \pi  x)
      +702026116366176 \cos (22 \pi  x) \nn\\
&& \phantom{\cF_{A,\p^7 J}^{(2)} =}
      +278039086913016 \cos (24 \pi  x)
      +46428148365984 \cos (26 \pi  x) \nn\\
&& \phantom{\cF_{A,\p^7 J}^{(2)} =}
      -90125157442320 \cos (28 \pi  x)
      -174503854036512 \cos (30 \pi  x) \nn\\
&& \phantom{\cF_{A,\p^7 J}^{(2)} =}
      +1714723915100625 \cos (32 \pi  x)],
\eea
\bea
&& \cF_{A,\p^8 J}^{(2)} = \f{1}{147573952589676412928}
[  98748495354471848514
 + 30498000650329745448\cos(2 \pi x) \nn\\
&& \phantom{\cF_{A,\p^8 J}^{(2)} =}
 + 7932922703252505486\cos(4 \pi x)
 + 3705192023982328224\cos(6 \pi x) \nn\\
&& \phantom{\cF_{A,\p^8 J}^{(2)} =}
 + 2200340128654313712\cos(8 \pi x)
 + 1488072097267195680\cos(10 \pi x) \nn\\
&& \phantom{\cF_{A,\p^8 J}^{(2)} =}
 + 1089248787499533252\cos(12 \pi x)
 + 836589062369563344\cos(14 \pi x) \nn\\
&& \phantom{\cF_{A,\p^8 J}^{(2)} =}
 + 649543136828812776\cos(16 \pi x)
 + 315395251414324048\cos(18 \pi x) \nn\\
&& \phantom{\cF_{A,\p^8 J}^{(2)} =}
 + 52944418339648308\cos(20 \pi x)
 + 21697904001447648\cos(22 \pi x) \nn\\
&& \phantom{\cF_{A,\p^8 J}^{(2)} =}
 + 10392917614189968\cos(24 \pi x)
 + 4877849135202912\cos(26 \pi x) \nn\\
&& \phantom{\cF_{A,\p^8 J}^{(2)} =}
 + 1791690648996105\cos(28 \pi x)
 - 75558258061572\cos(30 \pi x) \nn\\
&& \phantom{\cF_{A,\p^8 J}^{(2)} =}
 - 1262450781499050\cos(32 \pi x)
 - 2041674016882500\cos(34 \pi x) \nn\\
&& \phantom{\cF_{A,\p^8 J}^{(2)} =}
 + 21828296923200625\cos(36 \pi x) ].
\eea

\section{Results of Schatten distance in CFT}\label{appSchatten}

In this appendix, we collect the results of the second Schatten entropy between the RDMs of the ground and excited states of the current and its derivatives in the 2D free massless bosonic theory that are omitted in Section~\ref{secSchatten}.
We get
\be
D_2(\r_{A,G},\r_{A,J})^2 = \f{1}{256} [ 99 - 128 \cos(\pi x) + 28 \cos(2 \pi x) + \cos(4 \pi x) ],
\ee
\bea
&& D_2(\r_{A,G},\r_{A,\p J})^2 = \frac{1}{65536}
     [ 22931
     -28672 \cos (\pi  x)
     +8072 \cos (2 \pi  x)
     -4096 \cos (3 \pi  x) \nn\\
&& \phantom{D_2(\r_{A,G},\r_{A,\p J})^2 =}
     +1628 \cos (4 \pi  x)
     +56 \cos (6 \pi  x)
     +81 \cos (8 \pi  x) ], %\nn\\
\eea
\bea
&& D_2(\r_{A,G},\r_{A,\p^2J})^2 = \frac{1}{1048576}
      [358254
      -442368 \cos (\pi  x)
      +119016 \cos (2 \pi  x)
      -57344 \cos (3 \pi  x) \nn\\
&& \phantom{D_2(\r_{A,G},\r_{A,\p^2J})^2 =}
      +34431 \cos (4 \pi  x)
      -24576 \cos (5 \pi  x)
      +11044 \cos (6 \pi  x)
      +930 \cos (8 \pi  x) \nn\\
&& \phantom{D_2(\r_{A,G},\r_{A,\p^2J})^2 =}
      -12 \cos (10 \pi  x)
      +625 \cos (12 \pi  x)], %\nn\\
\eea
\bea
&& D_2(\r_{A,G},\r_{A,\p^3 J})^2 = \f{1}{4294967296}
      [1453496467
      -1784676352 \cos (\pi  x)
      +468409168 \cos (2 \pi  x) \nn\\
&& \phantom{D_2(\r_{A,G},\r_{A,\p^3 J})^2=}
      -220200960 \cos (3 \pi  x)
      +131718088 \cos (4 \pi  x)
      -90177536 \cos (5 \pi  x) \nn\\
&& \phantom{D_2(\r_{A,G},\r_{A,\p^3 J})^2=}
      +64487152 \cos (6 \pi  x)
      -52428800 \cos (7 \pi  x)
      +24720860 \cos (8 \pi  x) \nn\\
&& \phantom{D_2(\r_{A,G},\r_{A,\p^3 J})^2=}
      +2817488 \cos (10 \pi  x)
      +503800 \cos (12 \pi  x)
      -170000 \cos (14 \pi  x) \nn\\
&& \phantom{D_2(\r_{A,G},\r_{A,\p^3 J})^2=}
      +1500625 \cos (16 \pi  x)], %\nn\\
\eea
\bea
&& D_2(\r_{A,G},\r_{A,\p^4 J})^2 = \f{1}{68719476736}
      [23144240154
      -28332523520 \cos (\pi  x)
      +7332632360 \cos (2 \pi  x) \nn\\
&& \phantom{D_2(\r_{A,G},\r_{A,\p^4 J})^2 =}
      -3397386240 \cos (3 \pi  x)
      +2007130130 \cos (4 \pi  x)
      -1350565888 \cos (5 \pi  x) \nn\\
&& \phantom{D_2(\r_{A,G},\r_{A,\p^4 J})^2 =}
      +980954800 \cos (6 \pi  x)
      -765460480 \cos (7 \pi  x)
      +589623400 \cos (8 \pi  x) \nn\\
&& \phantom{D_2(\r_{A,G},\r_{A,\p^4 J})^2 =}
      -513802240 \cos (9 \pi  x)
      +248171248 \cos (10 \pi  x)
      +33050605 \cos (12 \pi  x) \nn\\
&& \phantom{D_2(\r_{A,G},\r_{A,\p^4 J})^2 =}
      +8911340 \cos (14 \pi  x)
      +1240190 \cos (16 \pi  x)
      -1968820 \cos (18 \pi  x) \nn\\
&& \phantom{D_2(\r_{A,G},\r_{A,\p^4 J})^2 =}
      +15752961 \cos (20 \pi  x)],
\eea
\bea
&& D_2(\r_{A,J},\r_{A,\p J})^2 =
      \frac{1}{65536}
      [21139
      -30720 \cos (\pi  x)
      +11400 \cos (2 \pi  x)
      -2048 \cos (3 \pi  x) \nn\\
&& \phantom{D_2(\r_{A,J},\r_{A,\p J})^2 =}
      +348 \cos (4 \pi  x)
      -200 \cos (6 \pi  x)
      +81 \cos (8 \pi  x)], %\nn\\
\eea
\bea
&& D_2(\r_{A,J},\r_{A,\p^2 J})^2 =
      \f{1}{1048576}
      [374606
      -483328 \cos (\pi  x)
      +117224 \cos (2 \pi  x)
      -28672 \cos (3 \pi  x)\nn\\
&& \phantom{D_2(\r_{A,J},\r_{A,\p^2 J})^2 =}
      +30719 \cos (4 \pi  x)
      -12288 \cos (5 \pi  x)
      +2596 \cos (6 \pi  x)
      -1470 \cos (8 \pi  x) \nn\\
&& \phantom{D_2(\r_{A,J},\r_{A,\p^2 J})^2 =}
      -12 \cos (10 \pi  x)
      +625 \cos (12 \pi  x)], %\nn\\
\eea
\bea
&& D_2(\r_{A,J},\r_{A,\p^3 J})^2 =
      \frac{1}{4294967296}
      [1561761939
      -1966080000 \cos (\pi  x)
      +455564112 \cos (2 \pi  x) \nn\\
&& \phantom{D_2(\r_{A,J},\r_{A,\p^3 J})^2 =}
      -110100480 \cos (3 \pi  x)
      +74046408 \cos (4 \pi  x)
      -45088768 \cos (5 \pi  x) \nn\\
&& \phantom{D_2(\r_{A,J},\r_{A,\p^3 J})^2 =}
      +50200304 \cos (6 \pi  x)
      -26214400 \cos (7 \pi  x)
      +7681500 \cos (8 \pi  x) \nn\\
&& \phantom{D_2(\r_{A,J},\r_{A,\p^3 J})^2 =}
      -3605040 \cos (10 \pi  x)
      +503800 \cos (12 \pi  x)
      -170000 \cos (14 \pi  x) \nn\\
&& \phantom{D_2(\r_{A,J},\r_{A,\p^3 J})^2 =}
      +1500625 \cos (16 \pi  x)], %\nn\\
\eea
\bea
&& D_2(\r_{A,J},\r_{A,\p^4 J})^2 =
      \frac{1}{68719476736}
      [25142105114
      -31346130944 \cos (\pi  x)
      +7103780648 \cos (2 \pi  x) \nn\\
&& \phantom{D_2(\r_{A,J},\r_{A,\p^4 J})^2 =}
      -1698693120 \cos (3 \pi  x)
      +1121509394 \cos (4 \pi  x)
      -675282944 \cos (5 \pi  x) \nn\\
&& \phantom{D_2(\r_{A,J},\r_{A,\p^4 J})^2 =}
      +464400048 \cos (6 \pi  x)
      -382730240 \cos (7 \pi  x)
      +455602280 \cos (8 \pi  x) \nn\\
&& \phantom{D_2(\r_{A,J},\r_{A,\p^4 J})^2 =}
      -256901120 \cos (9 \pi  x)
      +87608048 \cos (10 \pi  x)
      -39202835 \cos (12 \pi  x) \nn\\
&& \phantom{D_2(\r_{A,J},\r_{A,\p^4 J})^2 =}
      +8911340 \cos (14 \pi  x)
      +1240190 \cos (16 \pi  x)
      -1968820 \cos (18 \pi  x) \nn\\
&& \phantom{D_2(\r_{A,J},\r_{A,\p^4 J})^2 =}
      +15752961 \cos (20 \pi  x)],
\eea
\bea
&& D_2(\r_{A,\p J},\r_{A,\p^2 J})^2 =
      \frac{1}{1048576}
      [310302
      -450560 \cos (\pi  x)
      +179176 \cos (2 \pi  x)
      -61440 \cos (3 \pi  x) \nn\\
&& \phantom{D_2(\r_{A,\p J},\r_{A,\p^2 J})^2 =}
      +33343 \cos (4 \pi  x)
      -12288 \cos (5 \pi  x)
      +740 \cos (6 \pi  x)
      +1842 \cos (8 \pi  x) \nn\\
&& \phantom{D_2(\r_{A,\p J},\r_{A,\p^2 J})^2 =}
      -1740 \cos (10 \pi  x)
      +625 \cos (12 \pi  x)],%\nn\\
\eea
\bea
&& D_2(\r_{A,\p J},\r_{A,\p^3 J})^2 =
      \frac{1}{4294967296}
      [1425782931
      -1831862272 \cos (\pi  x)
      +505862992 \cos (2 \pi  x) \nn\\
&& \phantom{D_2(\r_{A,\p J},\r_{A,\p^3 J})^2 =}
      -244318208 \cos (3 \pi  x)
      +179530696 \cos (4 \pi  x)
      -45088768 \cos (5 \pi  x) \nn\\
&& \phantom{D_2(\r_{A,\p J},\r_{A,\p^3 J})^2 =}
      +27442928 \cos (6 \pi  x)
      -26214400 \cos (7 \pi  x)
      +9639388 \cos (8 \pi  x) \nn\\
&& \phantom{D_2(\r_{A,\p J},\r_{A,\p^3 J})^2 =}
      +2407888 \cos (10 \pi  x)
      -4513800 \cos (12 \pi  x)
      -170000 \cos (14 \pi  x) \nn\\
&& \phantom{D_2(\r_{A,\p J},\r_{A,\p^3 J})^2 =}
      +1500625 \cos (16 \pi  x)],%\nn\\
\eea
\bea
&& D_2(\r_{A,\p J},\r_{A,\p^4 J})^2 =
      \frac{1}{68719476736}
      [23252177946
      -29198647296 \cos (\pi  x)
      +7975163688 \cos (2 \pi  x) \nn\\
&& \phantom{D_2(\r_{A,\p J},\r_{A,\p^4 J})^2 =}
      -3846176768 \cos (3 \pi  x)
      +1906040850 \cos (4 \pi  x)
      -675282944 \cos (5 \pi  x) \nn\\
&& \phantom{D_2(\r_{A,\p J},\r_{A,\p^4 J})^2 =}
      +850325168 \cos (6 \pi  x)
      -382730240 \cos (7 \pi  x)
      +315486312 \cos (8 \pi  x) \nn\\
&& \phantom{D_2(\r_{A,\p J},\r_{A,\p^4 J})^2 =}
      -256901120 \cos (9 \pi  x)
      +64424688 \cos (10 \pi  x)
      +31903725 \cos (12 \pi  x) \nn\\
&& \phantom{D_2(\r_{A,\p J},\r_{A,\p^4 J})^2 =}
      -50808340 \cos (14 \pi  x)
      +1240190 \cos (16 \pi  x)
      -1968820 \cos (18 \pi  x) \nn\\
&& \phantom{D_2(\r_{A,\p J},\r_{A,\p^4 J})^2 =}
      +15752961 \cos (20 \pi  x)],
\eea
\bea
&& D_2(\r_{A,\p^2 J},\r_{A,\p^3 J})^2 =
      \frac{1}{4294967296}
      [1244436627
      -1798307840 \cos (\pi  x)
      +699061072 \cos (2 \pi  x) \nn\\
&& \phantom{D_2(\r_{A,\p^2 J},\r_{A,\p^3 J})^2 =}
      -227540992 \cos (3 \pi  x)
      +137370568 \cos (4 \pi  x)
      -95420416 \cos (5 \pi  x) \nn\\
&& \phantom{D_2(\r_{A,\p^2 J},\r_{A,\p^3 J})^2 =}
      +62973680 \cos (6 \pi  x)
      -26214400 \cos (7 \pi  x)
      +2962908 \cos (8 \pi  x) \nn\\
&& \phantom{D_2(\r_{A,\p^2 J},\r_{A,\p^3 J})^2 =}
      -182832 \cos (10 \pi  x)
      +3371000 \cos (12 \pi  x)
      -4010000 \cos (14 \pi  x) \nn\\
&& \phantom{D_2(\r_{A,\p^2 J},\r_{A,\p^3 J})^2 =}
      +1500625 \cos (16 \pi  x)],%\nn\\
\eea
\bea
&& D_2(\r_{A,\p^2 J},\r_{A,\p^4 J})^2 =
      \frac{1}{68719476736}
      [22445046650
      -28661776384 \cos (\pi  x)
      +7619426600 \cos (2 \pi  x) \nn\\
&& \phantom{D_2(\r_{A,\p^2 J},\r_{A,\p^4 J})^2 =}
      -3577741312 \cos (3 \pi  x)
      +3035422994 \cos (4 \pi  x)
      -1480589312 \cos (5 \pi  x) \nn\\
&& \phantom{D_2(\r_{A,\p^2 J},\r_{A,\p^4 J})^2 =}
      +903443632 \cos (6 \pi  x)
      -382730240 \cos (7 \pi  x)
      +287428584 \cos (8 \pi  x) \nn\\
&& \phantom{D_2(\r_{A,\p^2 J},\r_{A,\p^4 J})^2 =}
      -256901120 \cos (9 \pi  x)
      +37666032 \cos (10 \pi  x)
      +50104045 \cos (12 \pi  x) \nn\\
&& \phantom{D_2(\r_{A,\p^2 J},\r_{A,\p^4 J})^2 =}
      +13803500 \cos (14 \pi  x)
      -46387810 \cos (16 \pi  x)
      -1968820 \cos (18 \pi  x) \nn\\
&& \phantom{D_2(\r_{A,\p^2 J},\r_{A,\p^4 J})^2 =}
      +15752961 \cos (20 \pi  x)],
\eea
\bea
&& D_2(\r_{A,\p^3 J},\r_{A,\p^4 J})^2 =
      \frac{1}{68719476736}
      [19734065738
      -28443672576 \cos (\pi  x)
      +10928146728 \cos (2 \pi  x) \nn\\
&& \phantom{D_2(\r_{A,\p^3 J},\r_{A,\p^4 J})^2 =}
      -3460300800 \cos (3 \pi  x)
      +2057135250 \cos (4 \pi  x)
      -1396703232 \cos (5 \pi  x) \nn\\
&& \phantom{D_2(\r_{A,\p^3 J},\r_{A,\p^4 J})^2 =}
      +1023151024 \cos (6 \pi  x)
      -802160640 \cos (7 \pi  x)
      +578286120 \cos (8 \pi  x) \nn\\
&& \phantom{D_2(\r_{A,\p^3 J},\r_{A,\p^4 J})^2 =}
      -256901120 \cos (9 \pi  x)
      +34835952 \cos (10 \pi  x)
      -309395 \cos (12 \pi  x) \nn\\
&& \phantom{D_2(\r_{A,\p^3 J},\r_{A,\p^4 J})^2 =}
      -895380 \cos (14 \pi  x)
      +29954190 \cos (16 \pi  x)
      -40384820 \cos (18 \pi  x) \nn\\
&& \phantom{D_2(\r_{A,\p^3 J},\r_{A,\p^4 J})^2 =}
      +15752961 \cos (20 \pi  x)],
\eea
\bea
&& D_2(\r_{A, J},\r_{A,\bar J})^2 =
      \frac{1}{8} \sin ^4\Big(\frac{\pi  x}{2}\Big) [ 11+4 \cos (\pi  x)+\cos (2 \pi  x) ],%\nn\\
\eea
\bea
&& D_2(\r_{A, J},\r_{A,\bar\p\bar J})^2 =
      \frac{1}{65536}
      [24723
      -30720 \cos (\pi  x)
      +7048 \cos (2 \pi  x)
      -2048 \cos (3 \pi  x) \nn\\
&& \phantom{D_2(\r_{A, J},\r_{A,\bar\p\bar J})^2 =}
      +860 \cos (4 \pi  x)
      +56 \cos (6 \pi  x)
      +81 \cos (8 \pi  x)],%\nn\\
\eea
\bea
&& D_2(\r_{A, J},\r_{A,\bar\p^2\bar J})^2 =
      \frac{1}{1048576}
      [391022
      -483328 \cos (\pi  x)
      +108776 \cos (2 \pi  x)
      -28672 \cos (3 \pi  x) \nn\\
&& \phantom{D_2(\r_{A, J},\r_{A,\bar\p^2\bar J})^2 =}
      +18047 \cos (4 \pi  x)
      -12288 \cos (5 \pi  x)
      +4900 \cos (6 \pi  x)
      +930 \cos (8 \pi  x)\nn\\
&& \phantom{D_2(\r_{A, J},\r_{A,\bar\p^2\bar J})^2 =}
      -12 \cos (10 \pi  x)
      +625 \cos (12 \pi  x)],%\nn\\
\eea
\bea
&& D_2(\r_{A, J},\r_{A,\bar\p^3\bar J})^2 =
      \frac{1}{4294967296}
      [1594529939
      -1966080000 \cos (\pi  x)
      +436951888 \cos (2 \pi  x) \nn\\
&& \phantom{D_2(\r_{A, J},\r_{A,\bar\p^3\bar J})^2 =}
      -110100480 \cos (3 \pi  x)
      +70900680 \cos (4 \pi  x)
      -45088768 \cos (5 \pi  x) \nn\\
&& \phantom{D_2(\r_{A, J},\r_{A,\bar\p^3\bar J})^2 =}
      +28835568 \cos (6 \pi  x)
      -26214400 \cos (7 \pi  x)
      +11613660 \cos (8 \pi  x) \nn\\
&& \phantom{D_2(\r_{A, J},\r_{A,\bar\p^3\bar J})^2 =}
      +2817488 \cos (10 \pi  x)
      +503800 \cos (12 \pi  x)
      -170000 \cos (14 \pi  x) \nn\\
&& \phantom{D_2(\r_{A, J},\r_{A,\bar\p^3\bar J})^2 =}
      +1500625 \cos (16 \pi  x)],%\nn\\
\eea
\bea
&& D_2(\r_{A, J},\r_{A,\bar\p^4\bar J})^2 =
      \frac{1}{68719476736}
      [25456350234
      -31346130944 \cos (\pi  x)
      +6916347688 \cos (2 \pi  x) \nn\\
&& \phantom{D_2(\r_{A, J},\r_{A,\bar\p^4\bar J})^2 =}
      -1698693120 \cos (3 \pi  x)
      +1088577554 \cos (4 \pi  x)
      -675282944 \cos (5 \pi  x)\nn\\
&& \phantom{D_2(\r_{A, J},\r_{A,\bar\p^4\bar J})^2 =}
      +451948208 \cos (6 \pi  x)
      -382730240 \cos (7 \pi  x)
      +269807720 \cos (8 \pi  x)\nn\\
&& \phantom{D_2(\r_{A, J},\r_{A,\bar\p^4\bar J})^2 =}
      -256901120 \cos (9 \pi  x)
      +119720688 \cos (10 \pi  x)
      +33050605 \cos (12 \pi  x)\nn\\
&& \phantom{D_2(\r_{A, J},\r_{A,\bar\p^4\bar J})^2 =}
      +8911340 \cos (14 \pi  x)
      +1240190 \cos (16 \pi  x)
      -1968820 \cos (18 \pi  x)\nn\\
&& \phantom{D_2(\r_{A, J},\r_{A,\bar\p^4\bar J})^2 =}
      +15752961 \cos (20 \pi  x)],
\eea
\bea
&& D_2(\r_{A,\p J},\r_{A,\bar\p \bar J})^2 =
      \frac{1}{32768}
      [11539
      -14336 \cos (\pi  x)
      +4040 \cos (2 \pi  x)
      -2048 \cos (3 \pi  x) \nn\\
&& \phantom{D_2(\r_{A,\p J},\r_{A,\bar\p \bar J})^2 =}
      +732 \cos (4 \pi  x)
      -8 \cos (6 \pi  x)
      +81 \cos (8 \pi  x)],%\nn\\
\eea
\bea
&& D_2(\r_{A,\p J},\r_{A,\bar\p^2\bar J})^2 =
      \frac{1}{1048576}
      [364446
      -450560 \cos (\pi  x)
      +124264 \cos (2 \pi  x)
      -61440 \cos (3 \pi  x) \nn\\
&& \phantom{D_2(\r_{A,\p J},\r_{A,\bar\p^2\bar J})^2 =}
      +28735 \cos (4 \pi  x)
      -12288 \cos (5 \pi  x)
      +4772 \cos (6 \pi  x)
      +1458 \cos (8 \pi  x)\nn\\
&& \phantom{D_2(\r_{A,\p J},\r_{A,\bar\p^2\bar J})^2 =}
      -12 \cos (10 \pi  x)
      +625 \cos (12 \pi  x)],%\nn\\
\eea
\bea
&& D_2(\r_{A,\p J},\r_{A,\bar\p^3\bar J})^2 =
      \frac{1}{4294967296}
      [1485281427
      -1831862272 \cos (\pi  x)
      +500259664 \cos (2 \pi  x) \nn\\
&& \phantom{D_2(\r_{A,\p J},\r_{A,\bar\p^3\bar J})^2 =}
      -244318208 \cos (3 \pi  x)
      +113105864 \cos (4 \pi  x)
      -45088768 \cos (5 \pi  x)\nn\\
&& \phantom{D_2(\r_{A,\p J},\r_{A,\bar\p^3\bar J})^2 =}
      +30080752 \cos (6 \pi  x)
      -26214400 \cos (7 \pi  x)
      +15742428 \cos (8 \pi  x)\nn\\
&& \phantom{D_2(\r_{A,\p J},\r_{A,\bar\p^3\bar J})^2 =}
      +1179088 \cos (10 \pi  x)
      +503800 \cos (12 \pi  x)
      -170000 \cos (14 \pi  x)\nn\\
&& \phantom{D_2(\r_{A,\p J},\r_{A,\bar\p^3\bar J})^2 =}
      +1500625 \cos (16 \pi  x)],%\nn\\
\eea
\bea
&& D_2(\r_{A,\p J},\r_{A,\bar\p^4\bar J})^2 =
      \frac{1}{68719476736}
      [23705359386
      -29198647296 \cos (\pi  x)
      +7928223528 \cos (2 \pi  x) \nn\\
&& \phantom{D_2(\r_{A,\p J},\r_{A,\bar\p^4\bar J})^2 =}
      -3846176768 \cos (3 \pi  x)
      +1766285330 \cos (4 \pi  x)
      -675282944 \cos (5 \pi  x)\nn\\
&& \phantom{D_2(\r_{A,\p J},\r_{A,\bar\p^4\bar J})^2 =}
      +454569648 \cos (6 \pi  x)
      -382730240 \cos (7 \pi  x)
      +352514152 \cos (8 \pi  x)\nn\\
&& \phantom{D_2(\r_{A,\p J},\r_{A,\bar\p^4\bar J})^2 =}
      -256901120 \cos (9 \pi  x)
      +111856368 \cos (10 \pi  x)
      +16994285 \cos (12 \pi  x)\nn\\
&& \phantom{D_2(\r_{A,\p J},\r_{A,\bar\p^4\bar J})^2 =}
      +8911340 \cos (14 \pi  x)
      +1240190 \cos (16 \pi  x)
      -1968820 \cos (18 \pi  x)\nn\\
&& \phantom{D_2(\r_{A,\p J},\r_{A,\bar\p^4\bar J})^2 =}
      +15752961 \cos (20 \pi  x)],
\eea
\bea
&& D_2(\r_{A,\p^2 J},\r_{A,\bar\p^2\bar J})^2 =
      \frac{1}{524288}
      [179598
      -221184 \cos (\pi  x)
      +59592 \cos (2 \pi  x)
      -28672 \cos (3 \pi  x) \nn\\
&& \phantom{D_2(\r_{A,\p^2 J},\r_{A,\bar\p^2\bar J})^2 =}
      +17151 \cos (4 \pi  x)
      -12288 \cos (5 \pi  x)
      +5076 \cos (6 \pi  x)
      +258 \cos (8 \pi  x)\nn\\
&& \phantom{D_2(\r_{A,\p^2 J},\r_{A,\bar\p^2\bar J})^2 =}
      -156 \cos (10 \pi  x)
      +625 \cos (12 \pi  x)],%\nn\\
\eea
\bea
&& D_2(\r_{A,\p^2 J},\r_{A,\bar\p^3\bar J})^2 =
      \frac{1}{4294967296}
      [1463629971
      -1798307840 \cos (\pi  x)
      +478534480 \cos (2 \pi  x) \nn\\
&& \phantom{D_2(\r_{A,\p^2 J},\r_{A,\bar\p^3\bar J})^2 =}
      -227540992 \cos (3 \pi  x)
      +136129480 \cos (4 \pi  x)
      -95420416 \cos (5 \pi  x)\nn\\
&& \phantom{D_2(\r_{A,\p^2 J},\r_{A,\bar\p^3\bar J})^2 =}
      +52707056 \cos (6 \pi  x)
      -26214400 \cos (7 \pi  x)
      +12424668 \cos (8 \pi  x)\nn\\
&& \phantom{D_2(\r_{A,\p^2 J},\r_{A,\bar\p^3\bar J})^2 =}
      +277968 \cos (10 \pi  x)
      +2449400 \cos (12 \pi  x)
      -170000 \cos (14 \pi  x)\nn\\
&& \phantom{D_2(\r_{A,\p^2 J},\r_{A,\bar\p^3\bar J})^2 =}
      +1500625 \cos (16 \pi  x)],%\nn\\
\eea
\bea
&& D_2(\r_{A,\p^2 J},\r_{A,\bar\p^4\bar J})^2 =
      \frac{1}{68719476736}
      [23357789210
      -28661776384 \cos (\pi  x)
      +7579006760 \cos (2 \pi  x) \nn\\
&& \phantom{D_2(\r_{A,\p^2 J},\r_{A,\bar\p^4\bar J})^2 =}
      -3577741312 \cos (3 \pi  x)
      +2128388114 \cos (4 \pi  x)
      -1480589312 \cos (5 \pi  x)\nn\\
&& \phantom{D_2(\r_{A,\p^2 J},\r_{A,\bar\p^4\bar J})^2 =}
      +819416752 \cos (6 \pi  x)
      -382730240 \cos (7 \pi  x)
      +303984744 \cos (8 \pi  x)\nn\\
&& \phantom{D_2(\r_{A,\p^2 J},\r_{A,\bar\p^4\bar J})^2 =}
      -256901120 \cos (9 \pi  x)
      +102247152 \cos (10 \pi  x)
      +50991085 \cos (12 \pi  x)\nn\\
&& \phantom{D_2(\r_{A,\p^2 J},\r_{A,\bar\p^4\bar J})^2 =}
      +2890220 \cos (14 \pi  x)
      +1240190 \cos (16 \pi  x)
      -1968820 \cos (18 \pi  x)\nn\\
&& \phantom{D_2(\r_{A,\p^2 J},\r_{A,\bar\p^4\bar J})^2 =}
      +15752961 \cos (20 \pi  x)],
\eea
\bea
&& D_2(\r_{A,\p^3 J},\r_{A,\bar\p^3\bar J})^2 =
      \frac{1}{2147483648}
      [727774355
      -892338176 \cos (\pi  x)
      +234401872 \cos (2 \pi  x) \nn\\
&& \phantom{D_2(\r_{A,\p^3 J},\r_{A,\bar\p^3\bar J})^2 =}
      -110100480 \cos (3 \pi  x)
      +65888712 \cos (4 \pi  x)
      -45088768 \cos (5 \pi  x) \nn\\
&& \phantom{D_2(\r_{A,\p^3 J},\r_{A,\bar\p^3\bar J})^2 =}
      +32036336 \cos (6 \pi  x)
      -26214400 \cos (7 \pi  x)
      +11516380 \cos (8 \pi  x)\nn\\
&& \phantom{D_2(\r_{A,\p^3 J},\r_{A,\bar\p^3\bar J})^2 =}
      +1000144 \cos (10 \pi  x)
      -46600 \cos (12 \pi  x)
      -330000 \cos (14 \pi  x)\nn\\
&& \phantom{D_2(\r_{A,\p^3 J},\r_{A,\bar\p^3\bar J})^2 =}
      +1500625 \cos (16 \pi  x)],%\nn\\
\eea
\bea
&& D_2(\r_{A,\p^3 J},\r_{A,\bar\p^4\bar J})^2 =
      \frac{1}{68719476736}
      [23227903818
      -28443672576 \cos (\pi  x)
      +7418856488 \cos (2 \pi  x) \nn\\
&& \phantom{D_2(\r_{A,\p^3 J},\r_{A,\bar\p^4\bar J})^2 =}
      -3460300800 \cos (3 \pi  x)
      +2058681490 \cos (4 \pi  x)
      -1396703232 \cos (5 \pi  x)\nn\\
&& \phantom{D_2(\r_{A,\p^3 J},\r_{A,\bar\p^4\bar J})^2 =}
      +1002489264 \cos (6 \pi  x)
      -802160640 \cos (7 \pi  x)
      +476157480 \cos (8 \pi  x)\nn\\
&& \phantom{D_2(\r_{A,\p^3 J},\r_{A,\bar\p^4\bar J})^2 =}
      -256901120 \cos (9 \pi  x)
      +131964912 \cos (10 \pi  x)
      +11661165 \cos (12 \pi  x)\nn\\
&& \phantom{D_2(\r_{A,\p^3 J},\r_{A,\bar\p^4\bar J})^2 =}
      -3874580 \cos (14 \pi  x)
      +22114190 \cos (16 \pi  x)
      -1968820 \cos (18 \pi  x)\nn\\
&& \phantom{D_2(\r_{A,\p^3 J},\r_{A,\bar\p^4\bar J})^2 =}
      +15752961 \cos (20 \pi  x)],
\eea
\bea
&& D_2(\r_{A,\p^4 J},\r_{A,\bar\p^4\bar J})^2 =
      \frac{1}{34359738368}
      [11582265786
      -14166261760 \cos (\pi  x)
      +3668183880 \cos (2 \pi  x) \nn\\
&& \phantom{D_2(\r_{A,\p^4 J},\r_{A,\bar\p^4\bar J})^2 =}
      -1698693120 \cos (3 \pi  x)
      +1004387090 \cos (4 \pi  x)
      -675282944 \cos (5 \pi  x) \nn\\
&& \phantom{D_2(\r_{A,\p^4 J},\r_{A,\bar\p^4\bar J})^2 =}
      +490052720 \cos (6 \pi  x)
      -382730240 \cos (7 \pi  x)
      +292523240 \cos (8 \pi  x) \nn\\
&& \phantom{D_2(\r_{A,\p^4 J},\r_{A,\bar\p^4\bar J})^2 =}
      -256901120 \cos (9 \pi  x)
      +116695472 \cos (10 \pi  x)
      +12827885 \cos (12 \pi  x) \nn\\
&& \phantom{D_2(\r_{A,\p^4 J},\r_{A,\bar\p^4\bar J})^2 =}
      +1730780 \cos (14 \pi  x)
      -1621410 \cos (16 \pi  x)
      -2929220 \cos (18 \pi  x) \nn\\
&& \phantom{D_2(\r_{A,\p^4 J},\r_{A,\bar\p^4\bar J})^2 =}
      +15752961 \cos (20 \pi  x)].
\eea

%\bibliographystyle{D:/00.bibx/JHEPx}
%\bibliography{D:/00.bibx/2020,D:/00.bibx/2019,D:/00.bibx/2018,D:/00.bibx/1960,D:/00.bibx/1970,D:/00.bibx/1980,D:/00.bibx/1990,D:/00.bibx/1995,D:/00.bibx/1996,D:/00.bibx/1997,D:/00.bibx/1998,D:/00.bibx/1999,D:/00.bibx/2000,D:/00.bibx/2001,D:/00.bibx/2002,D:/00.bibx/2003,D:/00.bibx/2004,D:/00.bibx/2005,D:/00.bibx/2006,D:/00.bibx/2007,D:/00.bibx/2008,D:/00.bibx/2009,D:/00.bibx/2010,D:/00.bibx/2011,D:/00.bibx/2012,D:/00.bibx/2013,D:/00.bibx/2014,D:/00.bibx/2015,D:/00.bibx/2016,D:/00.bibx/2017,D:/00.bibx/book,D:/00.bibx/work,D:/00.bibx/thesis}

\providecommand{\href}[2]{#2}\begingroup\raggedright\endgroup

\end{document}